\documentclass[reprint,onecolumn,showpacs,amsmath,amssymb,aps]{revtex4}
\usepackage{graphicx}
\usepackage{dcolumn}
\usepackage{bm}
\usepackage{comment}

\usepackage{color}
\usepackage{mathtools}

\usepackage{epsfig}
\usepackage{amssymb}
\usepackage{amsmath}
\usepackage{bm}
\usepackage{color}
\usepackage[colorlinks,bookmarks=false,citecolor=blue,linkcolor=red,urlcolor=blue]{hyperref}

\usepackage{wrapfig}

\newcommand{\bs}{\boldsymbol}

\newcommand{\be}{\begin{equation}}
\newcommand{\ee}{\end{equation}}

\begin{document}

\title{Equilibration and GGE in interacting-to-free Quantum Quenches in dimensions $d>1$}

\author{Spyros Sotiriadis}
\email{sotiriad@sissa.it}
\affiliation{SISSA and INFN, Trieste - Italy}

\author{Gabriele Martelloni}
\email{gmartell@sissa.it}
\affiliation{SISSA and INFN, Trieste - Italy}

\begin{abstract}
Ground states of interacting QFTs are non-gaussian states, i.e. their connected $n$-point correlation functions do not vanish for $n>2$, in contrast to the free QFT case. We show that when the ground state of an interacting QFT evolves under a free \emph{massive} QFT  for a long time (a scenario that can be realised by a Quantum Quench), the connected correlation functions decay and all local physical observables equilibrate to values that are given by a gaussian density matrix that keeps memory \emph{only} of the \emph{two-point} initial correlation function. The argument hinges upon the fundamental physical principle of cluster decomposition, which is valid for the ground state of a general QFT. An analogous result was already known to be valid in the case of $d=1$ spatial dimensions, where it is a special case of the so-called Generalised Gibbs Ensemble (GGE) hypothesis, and we now generalise it to higher dimensions. Moreover in the case of \emph{massless} free evolution, despite the fact that the evolution may not lead to equilibration but unbounded increase of correlations with time instead, the GGE gives correctly the \emph{leading order} asymptotic behaviour of correlation functions in the thermodynamic and large time limit. The demonstration is performed in the context of bosonic relativistic QFT, but the arguments apply more generally.
\end{abstract}


\maketitle


\section{Introduction}
The nature of equilibration in quantum statistical physics \cite{revq,revq1,revq2} is still an open problem. Imagine that we prepare an isolated quantum system in the ground state of some hamiltonian and let it evolve unitarily under a different hamiltonian. This can be implemented by an instantaneous change of the hamiltonian of the system, a protocol known as Quantum Quench \cite{cc06}. Since the initial state is a pure state, the system will always remain in a pure state and its evolution will be periodic or quasi-periodic, i.e. if we wait long enough it will return to the initial state or as close to it as desired. In the contrary, if we consider subsystems of the whole system, since they are not isolated, they are described by a reduced density matrix, which may correspond to a statistical ensemble. The period of quantum recurrences typically diverges with the system size, so that considering first the thermodynamic limit (i.e. infinite system size) and then the large time limit \cite{CEF} it is possible that the system exhibits stationary behaviour, at the level of its subsystems and local physical observables. This behaviour however is not necessarily thermal; in fact one-dimensional integrable systems exhibit equilibration to non thermal ensembles \cite{rdyo07}. Progress in experimental techniques has made it possible to observe quantum non-equilibrium dynamics \cite{uc, kww06, tc07, tetal11, cetal12, getal11,shr12,rsb13,exp2} and the occurrence of stationary behaviour different from thermal at large times \cite{kww06,exp1}. Theoretical attention has mainly focused on the equilibration mechanism in one-dimensional integrable systems \cite{rdyo07,rdyo07b,fm-10,SFM,Mussardo13,cc06,c06a,c06b,c06c,c06,ir-11,cdeo08,cra10,bs08,mgs,ce13,fe13,fcec13,gge-new3,SotiriadisCalabrese,bse-14,delf14} (see Ref. \cite{revq,revq1,revq2} for general reviews) \footnote{In the context of this paper the number of dimensions $d$ refers to the number of \emph{spatial} dimensions; specifically by ``one-dimensional'' systems we mean systems in one spatial dimension.}. 

In general (although not without exceptions), for non-integrable extended quantum systems it is expected that the stationary behaviour can be described by a thermal ensemble (Gibbs or Canonical ensemble) \cite{rig09,deu,gme,rigsre,sre,ekmr-14}. This expectation is based on the statistical principle of entropy maximisation under the constraint of energy conservation. In this scenario equilibration is equivalent to standard thermalisation. If we consider integrable systems, on the other hand, the picture changes dramatically: the number of local conserved charges is infinite and, due to these extra constraints, thermalisation is prohibited. The obvious generalisation is to construct an ensemble that contains all infinite conserved quantities \cite{Jaynes}, but still being economic (in the sense that not all projections into energy eigenstates are to be included, as this would amount to keeping all unnecessary information about the initial state). In the seminal paper \cite{rdyo07} a Generalised Gibbs Ensemble (GGE) that includes all local conserved charges was proposed to describe the long time stationary behaviour of one-dimensional integrable systems. This conjecture has been proved correct in many cases, in most of which the evolution is performed under a free or effectively free theory, but also in special cases of quantum quenches in genuinely interacting integrable theories \cite{cc-07,cardy12,fm-10,CEF,bs-08,scc-09,rig09,r-09a,c06a,c06b,c06c,caz,f-13,eef-12,se-12,ccss-11,rs-12,CE-08,CE-10,CSC13a,fe-13,RS13,Mossel,Pozsgay11,ce-13,KCC-13,f-14,ck-14,DMV,GGE-CFT}. Even in cases where failure of the GGE in its original form was observed (some cases of interacting integrable models possessing bound-state eigenstates \cite{WDNBFRC,BWFDNVC,PMWKZT,POZ1,POZ2,ANDREI}), an alternative version of the GGE, sometimes called ``Quench Action'' ensemble \cite{ce13} or ``overlap Thermodynamic Bether Ansatz'' remains still valid \cite{PMWKZT,POZ2}, while the insertion of \emph{quasi-local} conserved charges \cite{PROSEN,Prosen2,Prosen3,Pereira,Panfil} into the definition of the original GGE has been suggested as a suitable modification. 

It is tempting to consider the generalisation of the GGE idea in higher spatial dimensions. An analogous generalisation of results for the transport properties of one-dimensional quantum systems to $d>1$ spatial dimensions has been recently performed \cite{CM2014,BOSONICDOYON}. Moreover new applications of the gauge-gravity duality to many-body physics \cite{Hartnoll,MG,myers,HolographicDoyon,AMADO} are also suggestive in this direction. Given that the GGE hypothesis refers to integrable models, extending its validity to higher dimensions would be restricted by the same restrictions imposed by integrability itself. According to the Coleman-Mandula theorem \cite{Coleman-Mandula} and its extensions \cite{CM2,CMgen}, integrability imposes strong constraints to QFTs in dimensions $d>1$ so that any integrable massive relativistic \emph{bosonic} field theory in $d>1$ must be free. Therefore we focus on the free case of the Klein-Gordon theory in $d>1$, where the GGE conjecture is plausible despite the dimensionality. More generally, in the context of Quantum Field Theory, the introduction of the AdS/CFT conjecture \cite{MALDACENA1} showed the first examples of seemingly integrable theories in spatial dimensions $d>1$ \cite{Beisert:2010jr}: ${\cal N}=4$ Super-Yang-Mills (SYM) and ${\cal N}=6$ Super-Chern-Simons theories \cite{MALDACENA2}, that are conformal and maximally supersymmetric, respectively in $d=3$ and $d=2$. The check of validity of the GGE hypothesis in such supersymmetric theories is left for future investigation. It should also be mentioned that the GGE may be useful in higher dimensions even for systems evolving under an interacting hamiltonian, not as an exact description of the large time limit (since there are no exact local conserved charges), but rather as an approximate description of pre-thermalisation, i.e. of the transient stage of the evolution that is expected to gradually lead to thermalisation \cite{NIC14,MSF14}. 

In the case of free evolution in one-dimensional bosonic models, validity of the GGE has been demonstrated in various examples of gaussian initial states (i.e. for free-to-free quantum quenches) or equivalently states with factorised charge product expectation values. In Ref.s~\cite{cra10,SotiriadisCalabrese} it was shown that, neither gaussianity, nor factorisation of charge products in the initial state are necessary for the validity of the GGE hypothesis in $d=1$. Instead, in Ref.~\cite{SotiriadisCalabrese} it was shown that the GGE conjecture is valid for a general interacting-to-free quantum quench in a one-dimensional bosonic system and the cluster decomposition principle, which is a fundamental physical requirement for the ground state of any physical hamiltonian, was identified as the sufficient and necessary property of the initial state that ensures the validity of the GGE. It is not a priori obvious whether dimensionality plays a crucial role in the above conclusion or not. In this paper we investigate this question. We find that in the case of massive free post-quench hamiltonian in $d=2$ or 3, the GGE conjecture is still valid, i.e. the system equilibrates and the GGE describes correctly its stationary behaviour at large times. In the case of massless free post-quench hamiltonian on the other hand, the system does not always equilibrate but instead its field correlations keep increasing unbounded with time. In particular, in $d=2$ the system never equilibrates, no matter what the initial state is, while in $d=3$ it may or may not equilibrate, depending on whether the initial correlations decay exponentially or algebraically with a sufficiently low exponent, respectively. When equilibration does occur, the GGE turns out to be valid. We also observe that even in the cases where the system does not equilibrate, the \emph{leading asymptotic} behaviour of long time correlations in the thermodynamic limit depends on only the two-point initial correlation function. All other information about the initial state contributes only subleading corrections in the large time and thermodynamic limit, although these corrections may in fact diverge with the system size or time (of course slower than the leading order expressions). The GGE predicts correctly only the leading order behaviour in this limit, so it is still applicable but in a weak sense. The economy property that the leading large time behaviour depends on only partial information about the initial state, which is characteristic of the GGE, seems therefore to be a more general feature of unitary evolution of extended quantum systems.

The paper is organised as follows:
In Section~\ref{sec:eq+GGE} we introduce the general concepts of equilibration in integrable models, the GGE hypothesis, the role of locality and of the double limit, thermodynamic limit and large time limit. Next we calculate the large time form of correlation functions after an interacting-to-free quantum quench and compare them with the GGE predictions. We first focus on the case of massive evolution in Section~\ref{sec:massive}, while in Section~\ref{sec:massless} we study separately the case of massless evolution. Conclusions and outlook are discussed in the last Section~\ref{sec:c+o}. 


\section{Equilibration and GGE}
\label{sec:eq+GGE}

\subsection{Definition of the model and quench protocol}
\label{sec:eq+GGE1}

We consider the Klein-Gordon field theory in Minkowski space-time with spatial dimension $d>1$ and defined in a large box of volume $L^d$ with periodic boundary conditions. This system is described by the hamiltonian 
\begin{align}
H& =\frac{1}{2}\int d^dx \; \left(\pi^2(\bs{x}) - \left(\nabla\phi(\bs{x})\right)^2 + m^2 \phi^2(\bs{x}) \right) \nonumber \\
&=\frac{1}{2}\sum_{\bs{k}}\left(\tilde{\pi}_{\bs{k}}\tilde{\pi}_{-\bs{k}} + E_{\bs{k}}^{2}\tilde{\phi}_{\bs{k}}\tilde{\phi}_{-\bs{k}}\right) \nonumber \\
& = \sum_{\bs{k}} E_{\bs{k}} a_{\bs{k}}^\dagger a_{\bs{k}} + \text{const}, \label{hamiltonian}
\end{align}
where $a^\dagger_{\bs{k}}, a_{\bs{k}}$ are the creation and annihilation operators and $E_{\bs{k}} = \sqrt{\bs{k}^2 +m^2}$ is the relativistic dispersion relation. For completeness, we report standard definitions for this theory that are used throughout the subsequent calculations in Appendix~\ref{app:-1}. 

We will eventually consider the thermodynamic limit $L\to\infty$, since equilibration of the large time expectation values of local observables is only possible after this limit has been taken (the evolution of observables of any finite system exhibits quantum recurrences, i.e. periodic or quasi periodic behaviour, as will be discussed in more detail below). We will first focus on the massive case $m\neq 0$. This hamiltonian will play the role of the post-quench hamiltonian in our problem.

The initial state $|\Omega_0\rangle$ is assumed to be the ground state of another hamiltonian, which is left arbitrary for the moment, except that it is assumed to be symmetric under translations and rotations in space (and therefore the initial state is also translationally and rotationally symmetric). We will later distinguish two cases: when the initial state is the ground state of another free hamiltonian (with a different dispersion relation e.g. a different mass) or the ground state of an interacting hamiltonian. In the first case the initial state is gaussian, while in the second case it is not. In both cases the evolution is under a gaussian hamiltonian. 

Gaussian states are defined by their characteristic property that their only non-vanishing connected correlation functions are the one- and two-point functions. All other multi-point correlation functions in such states can be expressed diagrammatically in terms of the latter. More explicitly, choosing the one-point function to be zero (which can be done by a simple redefinition of the field $\phi$ so as to subtract its expectation value on the state), the multi-point correlation functions can be written in terms of the two-point function by means of \emph{Wick's theorem}, i.e. by summing over all possible pair contractions. Examples of gaussian states are the ground states of free (quadratic) hamiltonians, but also their coherent and squeezed coherent states. Non-pure (mixed) gaussian states include thermal ensembles of free hamiltonians, as well as the GGE defined below, since all these can be expressed as exponentials of operators quadratic in the field $\phi$ and its canonically conjugate field $\pi$. It should be emphasised that gaussianity refers to the particular choice of fields whose correlation functions we are interested in. In the present problem, `gaussian initial state' means that the above property holds for the correlation functions of the local fields $\phi$ and $\pi$ which are also the one in terms of which the post-quench hamiltonian $H$ is free.

We will also assume that the initial state is symmetric under the transformation $\phi\to-\phi$. This means that the initial one-point function vanishes $\langle \phi(\bs{x}) \rangle_0\equiv\langle\Omega_0| \phi(\bs{x}) |\Omega_0\rangle = 0$. In the opposite case, which is valid for example when the pre-quench ground state exhibits spontaneous symmetry breaking, the one-point function develops a trivial oscillatory behaviour (Appendix~\ref{app:0}).

\subsection{The GGE conjecture, locality and extensivity aspects}
\label{sec:eq+GGE2}

According to the GGE conjecture, the expectation value of any local observable (or subsystem of the system) in the thermodynamic and large time limit is given by a statistical ensemble, described by the GGE density matrix
\be
\rho_{GGE} = \exp\left({-\sum_i \beta_i Q_i}\right), \label{gge1}
\ee
where $Q_i$ are the local conserved charges of the integrable system under evolution and each $\beta_i$ is the Lagrange multiplier associated to $Q_i$ accounting for the respective conservation law. The values of $\beta_i$ are determined by the self-consistency condition that the values of the charges in the GGE $\langle Q_i \rangle_{GGE}$ (being local observables themselves) are the same as their values in the initial state $\langle Q_i \rangle_{0}$
\be
\langle Q_i \rangle_{GGE} \equiv \frac{\text{Tr} \{ Q_i \, \rho_{GGE} \} }{\text{Tr} \{ \rho_{GGE} \}}= \langle\Omega_0|Q_i|\Omega_0\rangle \equiv \langle Q_i \rangle_{0}.
\ee
This is the only information about the initial state that is contained in the GGE.

The GGE in its original form is supposed to contain all local conserved charges $Q_i$ and only those \cite{rdyo07,CEF}; no other constant of motion is included \footnote{``Local conserved charges" precisely means that the charge operators are spatial integrals of one-point operators corresponding to the local charge densities.}. However it is typically written in terms of occupation number operators of all independent modes in which the integrable system can be decomposed, which are non-local operators. In Klein-Gordon theory these are the occupation number operators of each of the momentum modes $n(\bs{k})=a^\dagger_{\bs{k}} a_{\bs{k}}$ which are conserved operators and manifestly non-local. In terms of these operators the (non-normalised) GGE density matrix is
\begin{align}
&\rho_{GGE} = \exp\left({- \sum_{\bs{k}} \beta(\bs{k}) n(\bs{k})}\right), \nonumber \\ 
&\text{with } \langle n(\bs{k}) \rangle_{GGE} = \langle n(\bs{k}) \rangle_{0}. \label{gge2}
\end{align}
In free field theories, like in the present case, this is an equivalent alternative form because, the set of independent local charges is a linear combination of the occupation number operators, provided there are no convergence problems arising in the continuum limit \cite{Panfil}. In fact in most demonstrations of validity of the GGE conjecture, the latter is written in the form (\ref{gge2}). The locality requirement is related to the expectation that in a statistical ensemble the (generalised) ``internal energy'' $E_{GGE}$ should be an extensive quantity, in order to ensure the extensivity of thermodynamic quantities of macroscopic subsystems.

However in some cases of genuinely interacting integrable models, the GGE in its original form in terms of local charges (\ref{gge1}) has been proven to fail to describe the large time limit \cite{WDNBFRC,PMWKZT,BWFDNVC,POZ1,POZ2,ANDREI}, while in the form (\ref{gge2}) and taking into account all independent modes it is still correct. Moreover (\ref{gge2}) gives correct predictions even in cases where it is not possible to express it in terms of local charges and even if there do not exist local charges at all, like in the case of condensed matter models confined in external potential traps or models with long-range harmonic couplings \cite{SS8,SS12}. In such cases the GGE internal energy may be non-extensive. 

In the present problem the GGE internal energy corresponding to (\ref{gge2}) is
\be
E_{GGE} = -\log Z_{GGE} + S_{GGE}= {\sum_{\bs{k}}\beta(\bs{k}) \langle n(\bs{k}) \rangle_{GGE}} 
= - \sum_{\bs{k}} \frac{\partial \log  Z_{GGE} }{\partial \log \beta_{\bs{k}}}  \label{Fgge},
\ee
where $S_{GGE}$ is the GGE von Neumann entropy 
\be
S_{GGE} = -\text{Tr} \{\rho'_{GGE} \log \rho'_{GGE}\}
\ee
with
\be
\rho'_{GGE} \equiv \frac{\rho_{GGE}}{Z_{GGE}}, 
\ee
the normalised GGE density matrix and 
\be
 Z_{GGE} = \text{Tr} \rho_{GGE},
\ee
the GGE partition function. 
Expressing $ n(\bs{k})$ in terms of $\phi(\bs{x},t)$ and $\pi(\bs{x},t)=\dot\phi(\bs{x},t)$ fields and using the translational invariance of the initial state, we finally find that
\begin{align}
& E_{GGE} = \frac{L^d}{2} \int d^dr \; [D_1(\bs{r}) C_0^{(0,0)}(\bs{0},\bs{r})  
 + iD_2(\bs{r}) (C_0^{(0,1)}(\bs{0},\bs{r}) - C_0^{(1,0)}(\bs{0},\bs{r})) + D_3(\bs{r}) C_0^{(1,1)}(\bs{0},\bs{r}) ]
\end{align}
where
\begin{align}
C_0^{(0,0)}(\bs{x},\bs{y}) = \langle \phi(\bs{x}) \phi(\bs{y}) \rangle_0 , \\
C_0^{(0,1)}(\bs{x},\bs{y}) = \langle \phi(\bs{x}) \pi(\bs{y}) \rangle_0 , \\
C_0^{(1,0)}(\bs{x},\bs{y}) = \langle \pi(\bs{x}) \phi(\bs{y}) \rangle_0 , \\
C_0^{(1,1)}(\bs{x},\bs{y}) = \langle \pi(\bs{x}) \pi(\bs{y}) \rangle_0 ,
\end{align}
are the initial correlation functions and
\begin{align}
D_1(\bs{r}) & = \int d^dk \; e^{i\bs{k}\cdot\bs{r}} \beta(\bs{k}) E_{\bs{k}},  \\
D_2(\bs{r}) & = \int d^dk \; e^{i\bs{k}\cdot\bs{r}} \beta(\bs{k}), \\
D_3(\bs{r}) & = \int d^dk \; e^{i\bs{k}\cdot\bs{r}} \beta(\bs{k}) / E_{\bs{k}} . 
\end{align}
Therefore, if $\beta(\bs{k}),E_{\bs{k}}$ and $\langle n(\bs{k})\rangle_{GGE}$ are analytic functions in a neighbourhood of $k=0$ (and decay sufficiently fast at large $k$), then $D_i(\bs{r})$, $i=1,2,3$, all decay exponentially fast with the distance $r$ and the internal energy $E_{GGE}$ is extensive. Note that in the massless case $m=0$ the dispersion relation $E_{\bs{k}}=|\bs{k}|$ is not analytic at $k=0$, therefore the internal energy is a non-extensive quantity in this case. 

In any free field theory the GGE density matrix in the form (\ref{gge2}) is a gaussian ensemble with respect to the bosonic field $\phi$, i.e. the multi-point correlation functions of $\phi$ satisfy Wick's theorem \footnote{For clarity, we mention that by ``Wick's theorem" we precisely mean the property of gaussian states or ensembles that their multi-point correlation functions are equal to the sum of all possible contractions in terms of two-point functions.}. This is because it is a function of the momentum occupation number operators $n(\bs{k})$ which are quadratic in terms of the bosonic field and its canonical momentum. Therefore if the GGE conjecture is valid, it  equivalently means that the gaussian evolution erases the memory of non-gaussian initial field correlations, i.e. that connected correlation functions decay with time in the thermodynamic limit.

\subsection{Validity of the GGE and the cluster decomposition argument}
\label{sec:eq+GGE3}

\begin{figure}[h!]
\begin{center}
\includegraphics[width=.43\textwidth]{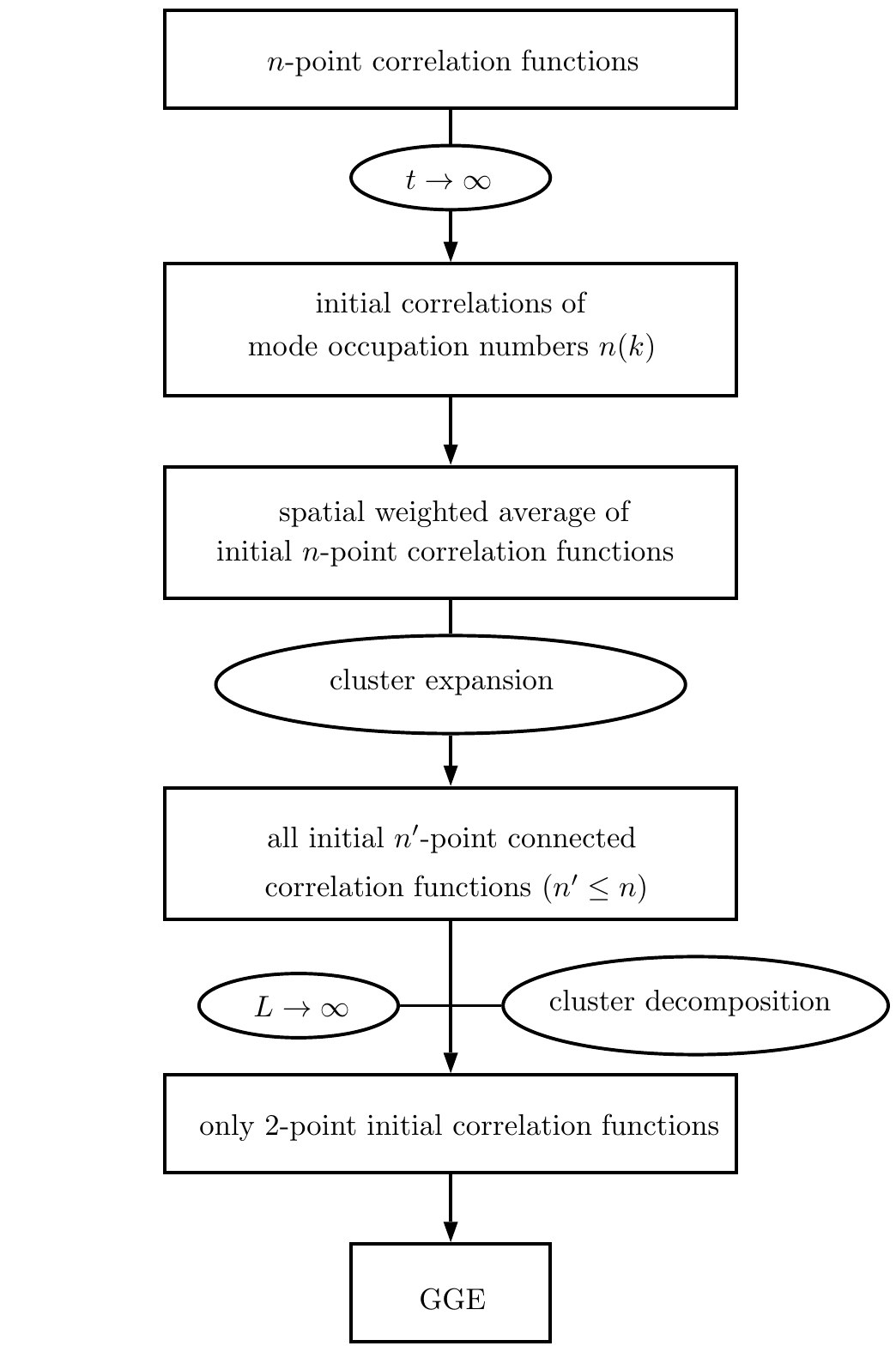}
\caption{Diagrammatic representation of the method. \label{fig:workflow}}
\end{center}
\end{figure}

We now outline our method to check the validity of the GGE conjecture based on the cluster decomposition principle. We have to calculate the evolution of local observables and, provided they equilibrate, to compare their large time stationary values with the GGE predictions. We will focus on the correlation functions of the bosonic field  $\phi(\bs{x},t)$. For the two-point function, testing the GGE is trivial, since, provided it equilibrates, its stationary value is given in terms of the conserved occupation numbers of the momentum modes $n(\bs{k})$, which by definition of the GGE are exactly the information contained in it. The first nontrivial test comes from the study of the four-point function, since its stationary value depends also on the initial \emph{correlations} of $n(\bs{k})$, i.e. on $\langle n(\bs{k}) n(\bs{p}) \rangle_{0}$ which are also conserved by the evolution but their values are not taken into account in the construction of the GGE. The GGE prediction for their values is simply the factorised value $\langle n(\bs{k})\rangle_{0} \langle n(\bs{p}) \rangle_{0}$, since it is a gaussian ensemble and Wick's theorem is valid in it. In the quantum quench case, if the pre-quench hamiltonian is also free, then Wick's theorem is valid in the initial state too and therefore $\langle n(\bs{k}) n(\bs{p}) \rangle_{0} = \langle n(\bs{k})\rangle_{0} \langle n(\bs{p}) \rangle_{0}$ and the GGE prediction is correct. If the pre-quench hamiltonian is not free, then $\langle n(\bs{k}) n(\bs{p}) \rangle_0$ does not factorise, however in the previously studied one-dimensional post-quench-massive case \cite{SotiriadisCalabrese} the GGE is still valid. The reason is that the contribution of connected correlations 
\be
\langle n(\bs{k}) n(\bs{p}) \rangle_0 - \langle n(\bs{k}) \rangle_0 \langle n(\bs{p}) \rangle_0
\ee 
to the large time value of the four-point function turns out to be a finite size correction, which vanishes in the thermodynamic limit $L\to\infty$. To see this we express the large time limit of correlation functions in terms of initial correlation functions and realise that they enter as a weighted spatial average over the whole system. The initial state after a quantum quench, being the ground state of some physical hamiltonian, satisfies the cluster decomposition principle, meaning that at large separations all of its multi-point correlation functions factorise to the fully contracted expression. As a consequence, when averaging them over the whole system, the only surviving information is the initial two-point correlation function, or equivalently the quantities $\langle n(\bs{k}) \rangle_0$, which information is precisely contained in the GGE. This argument works similarly for the large time limit of any other multi-point correlation function of order higher than four. The above reasoning is diagrammatically represented in the diagram of Fig.~\ref{fig:workflow}.

\subsection{Finite size, recurrences and the order of large time and thermodynamic limit}
\label{sec:eq+GGE4}

The order of the two limits $t\to\infty$ and $L\to\infty$ in the above method should be clarified. At finite volume the system exhibits recurrences (also called revivals), i.e. all correlation functions are periodic or quasi-periodic functions of time with a recurrence period that increases with $L$ \cite{revq1}. This is because the energy spectrum of a finite system is discrete with the spacing between two successive energy levels being of order $1/L$, due to the quantisation of momenta. If the energy eigenvalues are commensurate, i.e. they are all integer multiples of a fundamental frequency, then the evolution is exactly periodic. This is the case in $1d$ massless relativistic systems and the recurrence period is equal to the system size $L$, since the dispersion relation is $E_{k}=|k|=2\pi |n|/L$, with $n$ integer. If instead the energy eigenvalues are incommensurate, then revivals are partial and the evolution is quasi-periodic, meaning that the system will return as close to the initial state as we want, if we wait long enough for a sufficiently approximate revival. 

From the above we see that the limit ${t\to\infty}$ does \emph{not} exist at finite $L$. Instead, if the thermodynamic limit $L\to\infty$ is taken first, then the evolution is no longer periodic (the recurrence period diverges) and it \emph{is} now possible for the system and its correlation functions to exhibit stationary behaviour as $t\to\infty$. This means that the two limits do not commute
\be
\lim_{t\to\infty} \lim_{L\to\infty} \neq \lim_{L\to\infty} \lim_{t\to\infty},
\ee
as one may exist while the other does not. 
However let us consider the ``long time average" values of correlation functions at finite volume $L$, defined as
\be
\bar{C}^{(n)}(\bs{x}_1,...,\bs{x}_n) \equiv \lim_{T\to\infty} T^{-1} \int^T dt \, C_t^{(n)}(\bs{x}_1,...,\bs{x}_n),
\ee
where $C_t^{(n)}$ is the correlation function at time $t$ (and at finite volume). In fact if the evolution is strictly periodic (not quasi-periodic), then it is sufficient to average over a single period, which as mentioned above is an increasing function of $L$. Obviously the long time average values depend explicitly on $L$. If the system becomes stationary in the limit $\lim_{t\to\infty} \lim_{L\to\infty}$, then the stationary values of correlation functions are identical to the thermodynamic limit of the long time averaged values, i.e. 
\be
\lim_{L\to\infty} \lim_{T\to\infty} T^{-1} \int^T dt \; C_t^{(n)}(\bs{x}_1,...,\bs{x}_n) = \lim_{t\to\infty} \lim_{L\to\infty} C_t^{(n)}(\bs{x}_1,...,\bs{x}_n) \quad (\text{if equilibrated}).
\ee
Therefore the two limits can be effectively permuted if we employ time averaging. This has the technical advantage that long time averages in finite systems can be calculated easier. 
\begin{figure}[t]
\begin{center}
\includegraphics[width=.4\textwidth]{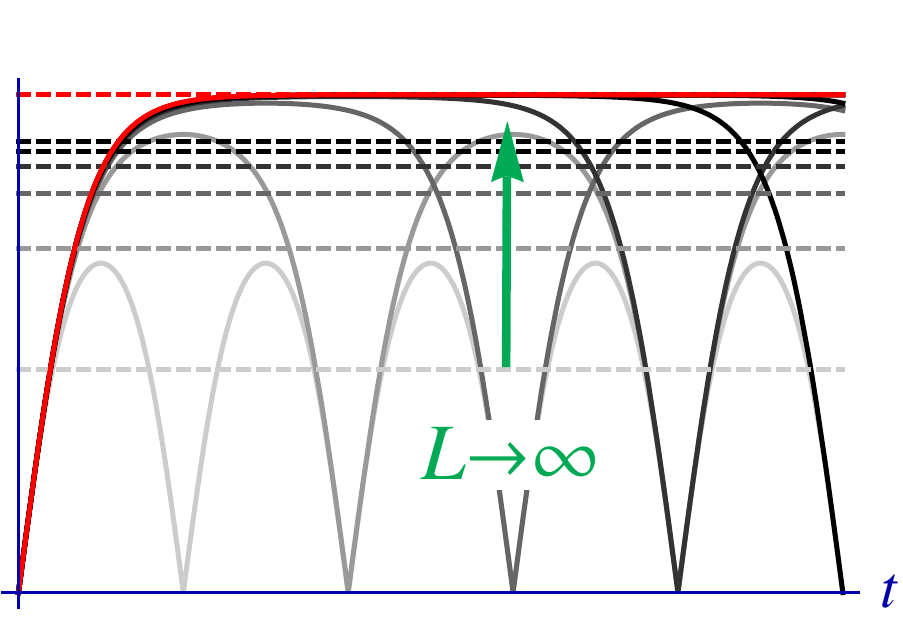}
\hspace{30pt}
\includegraphics[width=.4\textwidth]{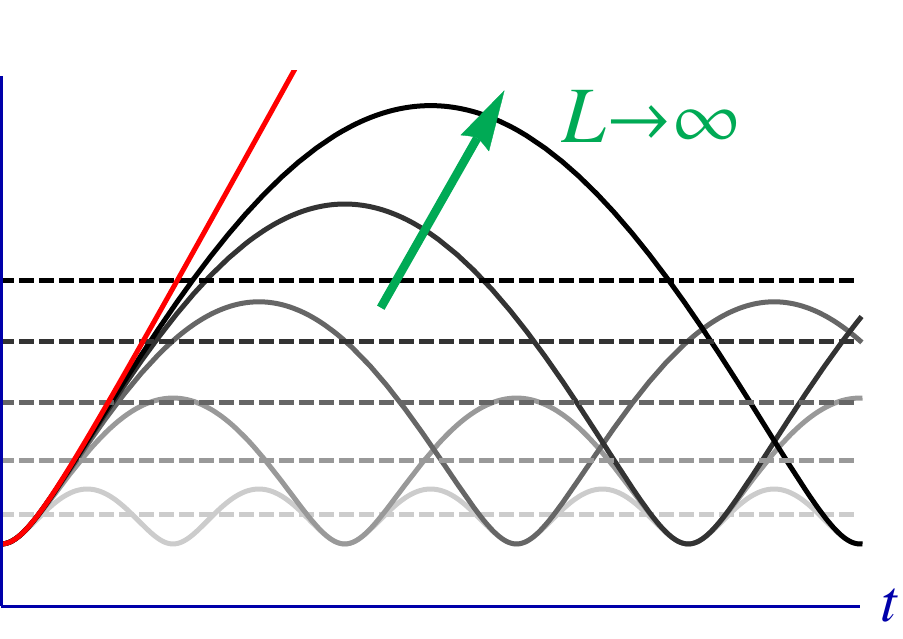}
\caption{Recurrences and scaling of long time averages: Typical plots of the time evolution of an observable (grey full lines) and its long time average value (dashed lines) for system sizes $L$ that increase successively by a factor of two (darker grey lines correspond to larger sizes $L$), along with their asymptotic behaviour in the thermodynamic limit $L\to\infty$ (red full and dashed lines). \emph{Top}: the case of equilibration. \emph{Bottom}: the case of unbounded increase. Notice that in the first case the time averages converge as $L\to\infty$ to the stationary value, while in the second case the unbounded increase with the time $t$ in the thermodynamic limit is reflected in the finite size system as unbounded scaling of the time averages with increasing $L$.}
\label{fig:evol}
\end{center}
\end{figure}

If, instead of strict equilibration, the system exhibits bounded persistent oscillations in the limit $\lim_{t\to\infty} \lim_{L\to\infty}$, these will obviously be lost after time averaging. If on the other hand there is neither equilibration nor equilibration on average, but instead correlations keep increasing indefinitely for large times, then the long time average will be divergent with $L$ in the thermodynamic limit (Fig.~\ref{fig:evol}). In general, equilibration means that the evolution of observables in finite size systems exhibits a ``plateau'' (typically accompanied by rapid finite size oscillations) during a long time window between two recurrences, that becomes infinite in the thermodynamic limit \cite{revq1,revq2}. Instead, if the system does not exhibit such a plateau, then the time averaged observables may increase, when the system size (and therefore the recurrence period) increases, and finally diverge in the thermodynamic limit. We can therefore study the scaling behaviour of the time averaged correlation functions as functions of $L$, in order to determine the scaling behaviour of the correlation functions as functions of time in the thermodynamic limit and determine whether they equilibrate or not.

\section{Massive non-interacting post-quench hamiltonian}
\label{sec:massive}

Having outlined our method and clarified certain important points, we are now ready to check the validity of the GGE, first focusing on the case of massive post-quench hamiltonian. We will therefore compare the long time averages of correlation functions of the field $\phi(\bs{x},t)$ with the GGE predictions. The calculation is presented in full detail in the Appendices. For the two-point correlation function $C_{t}^{(2)}(\bs{x}_1,\bs{x}_2)\equiv\langle\phi(\bs{x}_1;t)\phi(\bs{x}_2;t)\rangle$ we find that its long time average value is (Appendix~\ref{app:1})
\begin{align}
 & \bar{C}^{(2)}(\bs{x}_1,\bs{x}_2) = \bar{C}^{(2)}(\bs{x}_1-\bs{x}_2)  \nonumber \\ 
 & = \int{\frac{d^dk}{(2\pi)^d}  \frac{1}{2E_{k}}e^{i\bs{k}\cdot(\bs{x}_1-\bs{x}_2)}{(1+2 \langle n({\bs{k}})\rangle_{0} )}}.
\label{eq:Cinfty}
\end{align}
i.e. it is expressed in terms of  the momentum distribution $\langle n({\bs{k}})\rangle_{0}$ in the initial state, which is conserved under the evolution. The above expression can be written in terms of coordinate space initial correlation functions as
\begin{align}
\bar{C}^{(2)}(\bs{r}) & =\frac{1}{2}\bigg(C_{0}^{(2)}(\bs{r})+\int d^d s\, H(\bs{s})\ddot{C}_{0}^{(2)}(\bs{r}-\bs{s})\bigg),
\label{eq:Cinfty2}
\end{align}
where we defined 
\begin{align}
H(\bs{r})\equiv & \int\frac{d^dk}{(2\pi)^d}\frac{e^{i\bs{k}\cdot \bs{r}}}{E_{k}^{2}},\label{eq:Ha}
\end{align}
and 
\begin{align}
& \ddot{C}_{0}^{(2)}(\bs{r})  \equiv\left\langle \pi(\bs{r},0)\pi(\bs{0},0)\right\rangle \nonumber \\ 
 & =\left.\frac{\partial}{\partial t_{1}}\frac{\partial}{\partial t_{2}}\left\langle \phi(\bs{r},t_{1})\phi(\bs{0},t_{2})\right\rangle \right|_{t_{1}=t_{2}=0}.\label{eq:ddC}
\end{align}

This should be compared to the GGE prediction, which is (Appendix~\ref{app:3}) 
\begin{align}
& C_{GGE}^{(2)}(\bs{x}_1-\bs{x}_2) \nonumber \\
& = \int{\frac{d^dk}{(2\pi)^d}  \frac{1}{2E_{k}}e^{i\bs{k}\cdot(\bs{x}_1-\bs{x}_2)}{(1+2 \langle n({\bs{k}})\rangle_{GGE} )}}.
\label{GGE_2pf}
\end{align}
i.e. it is expressed in terms of  the momentum distribution in the GGE $\langle n({\bs{k}})\rangle_{GGE}$. The two expressions (\ref{eq:Cinfty}) and (\ref{GGE_2pf}) are trivially identical, i.e.
\begin{equation}
\bar{C}^{(2)}(\bs{r}) = C_{GGE}^{(2)}(\bs{r}) \label{eq:GGEvalid}
\end{equation}
due to the defining property of the GGE that 
\begin{equation}
\langle n(\bs{k}) \rangle_{GGE} = \langle n(\bs{k}) \rangle_{0}.
\end{equation}

Next, for the four-point correlation function $C_{t}^{(4)}(\bs{x}_{1},\bs{x}_{2},\bs{x}_{3},\bs{x}_{4}) \equiv \langle\phi(\bs{x}_{1};t)\phi(\bs{x}_{2};t)\phi(\bs{x}_{3};t)\phi(\bs{x}_{4};t)\rangle$, the long time average value is (Appendix~\ref{app:2}) 
\begin{align}
 & \bar{C}^{(4)}(\bs{x}_{1},\bs{x}_{2},\bs{x}_{3},\bs{x}_{4}) \nonumber \\
 & =\frac{1}{2}\int\frac{d^d{k}\, d^d{p}}{(2\pi)^{2d}}\frac{1}{4E_{{k}}E_{{p}}}S(\bs{k},\bs{p};\bs{x}_{1},\bs{x}_{2},\bs{x}_{3},\bs{x}_{4})  \nonumber \\
 & \times \left(\langle n({\bs{k}})n({\bs{p}})\rangle_{0}+\langle n({\bs{k}})\rangle_{0}+\frac{1}{4}\right),\label{eq:lt2a}
\end{align}
where 
\be
S(\bs{k},\bs{p};\bs{x}_{1},\bs{x}_{2},\bs{x}_{3},\bs{x}_{4})\equiv\sum_{{\text{all perm.s}\atop \text{of 1,2,3,4}}}e^{i\bs{k}\cdot(\bs{x}_{2}-\bs{x}_{1})+i\bs{p}\cdot(\bs{x}_{4}-\bs{x}_{3})}.
\label{S-function}
\ee
In terms of initial correlation functions in coordinate space, the above can be written as
\begin{align}
& \bar{C}^{(4)}(\bs{x}_{1},\bs{x}_{2},\bs{x}_{3},\bs{x}_{4}) = \nonumber \\
& \frac{1}{32}\sum_{{\text{all perm.s}\atop \text{of } 1,2,3,4}}  \iint\limits_{L^d} \frac{d^dsd^dr }{L^{2d}} \, 
\Bigg [ C_{0}^{(4)} \left(\bs{s},\bs{s}+\bs{x}_{1}-\bs{x}_{2},\bs{r},\bs{r}+\bs{x}_{3}-\bs{x}_{4}\right) \nonumber \\
& + \int\limits_{L^d}d^ds' \, H(\bs{s}') \ddot{C}_{0}^{(4)} (\bs{s},\bs{s}'+\bs{s}+\bs{x}_{1}-\bs{x}_{2},\bs{r},\bs{r}+\bs{x}_{3}-\bs{x}_{4}) \nonumber \\
& + \int\limits_{L^d}d^dr' \, H(\bs{r}') \big(\ddot{C}_{0}^{(4)}(\bs{s},\bs{s}+\bs{x}_{1}-\bs{x}_{2},\bs{r},\bs{r}'+\bs{r}+\bs{x}_{3}-\bs{x}_{4})\big)^{*}  \nonumber \\
& + \iint\limits_{L^d}d^ds'd^dr' \, H(\bs{s}')H(\bs{r}') 
\ddddot{C}_{0}^{(4)}(\bs{s},\bs{s}'+\bs{s}+\bs{x}_{1}-\bs{x}_{2},\bs{r},\bs{r}'+\bs{r}+\bs{x}_{3}-\bs{x}_{4})  \Bigg ], \label{eq:6a}
\end{align}
where 
\be
\ddot{C}_{0}^{(4)}(\bs{x}_1,\bs{x}_2,\bs{x}_3,\bs{x}_4) = \left\langle \pi(\bs{x}_{1})\pi(\bs{x}_{2})\phi(\bs{x}_{3})\phi(\bs{x}_{4})\right\rangle _{0} \label{ddC4}
\ee
and 
\be
\ddddot{C}_{0}^{(4)}(\bs{x}_1,\bs{x}_2,\bs{x}_3,\bs{x}_4) = \left\langle \pi(\bs{x}_{1})\pi(\bs{x}_{2})\pi(\bs{x}_{3})\pi(\bs{x}_{4})\right\rangle _{0} \label{ddddC4}
\ee
while the function $H(\bs{x})$ has been defined in (\ref{eq:Ha}). The last relation shows that \emph{the long time average of the four-point function $\bar{C}^{(4)}$ can be expressed as a weighted spatial average of initial correlation functions $C_{0}^{(4)},\ddot{C}_{0}^{(4)}$ and $\ddddot{C}_{0}^{(4)}$.}

Notice that (\ref{eq:lt2a}) depends not only on the initial momentum occupation numbers, but also \emph{on their initial correlations $\langle n({\bs{k}})n({\bs{p}})\rangle_{0}$}. In contrast, the GGE prediction is given in terms of solely the momentum occupation numbers $\langle n({\bs{k}})\rangle_{0}$. Indeed, since the GGE for a free model is a gaussian ensemble, Wick's theorem applies and therefore the correlations of the occupation numbers factorise 
\begin{equation}
\langle n({\bs{k}})n({\bs{p}})\rangle_{GGE}=\langle n({\bs{k}})\rangle_{GGE} \langle n({\bs{p}})\rangle_{GGE},
\end{equation}
thus leading to the prediction (Appendix~\ref{app:3})
\begin{align}
 & C_{GGE}^{(4)}(\bs{x}_{1},\bs{x}_{2},\bs{x}_{3},\bs{x}_{4}) \nonumber \\
 & =\frac{1}{2}\int\frac{d^d{k}\, d^d{p}}{(2\pi)^{2}}\frac{1}{4E_{\bs{k}}E_{\bs{p}}}S(\bs{k},\bs{p};\bs{x}_{1},\bs{x}_{2},\bs{x}_{3},\bs{x}_{4}) \nonumber \\
 & \times \left(\langle n({\bs{k}}) \rangle_{GGE} \langle n({\bs{p}})\rangle_{GGE}+\langle n({\bs{k}})\rangle_{GGE}+\frac{1}{4}\right),\label{eq:lt2ggea}
\end{align}
or simpler, in coordinate space
\begin{align}
 & C_{\text{GGE}}^{(4)}(\bs{x}_{1},\bs{x}_{2},\bs{x}_{3},\bs{x}_{4})=C_{\text{GGE}}^{(2)}(\bs{x}_{1},\bs{x}_{2})C_{\text{GGE}}^{(2)}(\bs{x}_{3},\bs{x}_{4}) \nonumber \\
 & +[2\leftrightarrow3]+[2\leftrightarrow4] . 
 \label{eq:G4ggea}
\end{align}

\subsection{Non-interacting pre-quench hamiltonian}

From the comparison of (\ref{eq:lt2a}) and (\ref{eq:lt2ggea}), an obvious case for which the GGE prediction is valid, is when the correlations of charges factorise also in the initial state
\begin{equation}
\langle n({\bs{k}})n({\bs{p}})\rangle_{0}=\langle n({\bs{k}})\rangle_{0} \langle n({\bs{p}})\rangle_{0},
\end{equation}
i.e. when Wick's theorem is valid in the initial state, as is the case of a ground state of a pre-quench hamiltonian that is also free (i.e. quadratic) in terms of the bosonic field $\phi$, though with different physical parameters (e.g. different mass). Therefore \emph{for a free-to-free quantum quench the GGE conjecture is always valid.} This is because the initial state is then gaussian, like the GGE itself, and therefore the consistency of the two-point function expressions (\ref{eq:Cinfty}) and (\ref{GGE_2pf}) is sufficient to ensure consistency for the four-point function and, following the same reasoning, for all multi-point correlation functions.

\subsection{Interacting pre-quench hamiltonian}

We now turn to the more interesting case of non-gaussian initial states, such as the ground states of interacting hamiltonians. We will show that, even though the above factorisation does not hold, the spatial averages of the initial four-point correlation function entering in (\ref{eq:6a}) are equal to the product of two-point correlation functions so that the GGE prediction is still correct.

Let us focus on the first term in the integral of (\ref{eq:6a}), which is essentially a spatial average of the initial four-point correlation function $C_{0}^{(4)}$ with respect to two coordinate variables $\bs{s}$ and $\bs{r}$. From the general cluster expansion for the $C_{0}^{(4)}$ correlation function, taking into account the translational invariance of the initial state and our assumption that $\langle\phi(\bs{x})\rangle_0=\phi_0$, we have
\begin{align}
 C_{0}^{(4)} & \left(\bs{s},\bs{s}+\bs{x}_{1}-\bs{x}_{2},\bs{r},\bs{r}+\bs{x}_{3}-\bs{x}_{4}\right)   \nonumber \\
 & = C_{0}^{(2)}(\bs{x}_{1}-\bs{x}_{2})C_{0}^{(2)}(\bs{x}_{3}-\bs{x}_{4}) \nonumber \\
 & + C_{0}^{(2)}(\bs{r}-\bs{s})C_{0}^{(2)}(\bs{r}-\bs{s}+\bs{x}_{3}-\bs{x}_{4}+\bs{x}_{2}-\bs{x}_{1}) \nonumber \\
 & + C_{0}^{(2)}(\bs{r}-\bs{s}+\bs{x}_{3}-\bs{x}_{4})C_{0}^{(2)}(\bs{r}-\bs{s}+\bs{x}_{2}-\bs{x}_{1}) \nonumber \\
 & + C_{0,c}^{(4)}(\bs{0},\bs{x}_{1}-\bs{x}_{2},\bs{r}-\bs{s},\bs{r}-\bs{s}+\bs{x}_{3}-\bs{x}_{4})
\end{align}
We then see that substituting into the integral, only the first term in the above expansion, which is independent of $\bs{s}$ and $\bs{r}$, survives. Indeed, due to the cluster decomposition principle all other terms give contributions that vanish in the thermodynamic limit. We therefore have
\begin{align}
 & \lim_{L\to\infty} \iint\limits_{L^{d}} \frac{d^dsd^dr }{L^{2d}} \, C_{0}^{(4)}\left(\bs{s},\bs{s}+\bs{x}_{1}-\bs{x}_{2},\bs{r},\bs{r}+\bs{x}_{3}-\bs{x}_{4}\right) \nonumber \\
 & =  C_{0}^{(2)}(\bs{x}_{1}-\bs{x}_{2}) C_{0}^{(2)}(\bs{x}_{3}-\bs{x}_{4}).
\end{align}
As shown in more detail in Appendix~\ref{app:4}, similar results hold for all other terms in the integral of (\ref{eq:6a}). For the last term, for example, we find 
\begin{align}
 & \lim_{L\to\infty} \iint\limits_{L^{d}} \frac{d^dsd^dr }{L^{2d}} \iint\limits_{L^d} d^ds'd^dr' \, H(\bs{s}')H(\bs{r}') \ddddot{C}_{0}^{(4)}(\bs{s},\bs{s}'+\bs{s}+\bs{x}_{1}-\bs{x}_{2},\bs{r},\bs{r}'+\bs{r}+\bs{x}_{3}-\bs{x}_{4}) \nonumber \\
  & = \iint d^ds' d^dr' \, H(\bs{s}') H(\bs{r}') \, \ddot{C}_{0}^{(2)}(\bs{x}_{1}-\bs{x}_{2}+\bs{s}') \ddot{C}_{0}^{(2)}(\bs{x}_{3}-\bs{x}_{4}+\bs{r}'),
\end{align}
where the last expression is always finite as $L\to\infty$ for a massive post-quench hamiltonian, since the function $H(\bs{r})$ decays exponentially with the distance $r$. 

Summing up all surviving terms, (\ref{eq:6a}) reduces to
\begin{align}
& \bar{C}^{(4)}(\bs{x}_{1},\bs{x}_{2},\bs{x}_{3},\bs{x}_{4})  = 
\nonumber \\ & = \frac{1}{32} \sum_{{\text{all perm.s}\atop \text{of }1,2,3,4}} \left(C_{0}^{(2)}(\bs{x}_{1}-\bs{x}_{2})+\int d^ds'H(\bs{s}')\ddot{C}_{0}^{(2)}(\bs{x}_{1}-\bs{x}_{2}+\bs{s}')\right) \nonumber \\
 & \times\left(C_{0}^{(2)}(\bs{x}_{3}-\bs{x}_{4})+\int d^dr'H(\bs{r}')\ddot{C}_{0}^{(2)}(\bs{x}_{3}-\bs{x}_{4}+\bs{r}')\right), \label{eq:C4a}
\end{align}
which, using our earlier result (\ref{eq:Cinfty2}) for the two-point correlation
function, can be identified with
\begin{align}
& \bar{C}^{(4)}(\bs{x}_{1},\bs{x}_{2},\bs{x}_{3},\bs{x}_{4}) \nonumber \\
 &  =  \frac{1}{8} \sum_{{\text{all perm.s}\atop \text{of }1,2,3,4}} 
 \bar{C}^{(2)}(\bs{x}_{1}-\bs{x}_{2})\bar{C}^{(2)}(\bs{x}_{3}-\bs{x}_{4}) \nonumber \\ 
& =\bar{C}^{(2)}(\bs{x}_{1}-\bs{x}_{2})\bar{C}^{(2)}(\bs{x}_{3}-\bs{x}_{4})+[2\leftrightarrow3]+[2\leftrightarrow4].\label{eq:C4ba}
\end{align}
In this last form it is manifest that Wick's theorem is valid for the long time average of the four-point function. Therefore comparing with the GGE expression (\ref{eq:G4ggea}) and since from (\ref{eq:GGEvalid}) the GGE prediction for the two-point function is correct, we conclude that the GGE is correct also for the four-point function
\be
\bar{C}^{(4)}(\bs{x}_{1},\bs{x}_{2},\bs{x}_{3},\bs{x}_{4}) = {C}_{\text{GGE}}^{(4)}(\bs{x}_{1},\bs{x}_{2},\bs{x}_{3},\bs{x}_{4}),
\ee
always assuming that the post-quench hamiltonian is massive.

Overall this means that, even when the initial state is a non-gaussian state, the contribution of connected initial correlations of the occupation number operators turns out to vanish in the thermodynamic limit, \emph{as a consequence of the cluster decomposition principle}. Therefore as anticipated, the GGE conjecture is still valid in this case.

\subsection{Higher order correlation functions}
\label{sec:hcf}

Higher order multi-point correlation functions can be calculated similarly and the above conclusions about equilibration and validity of the GGE prediction hold the same. Indeed the stationary or long time average value of the $2n$-point function $\bar{C}^{(2n)}(\left\{ \bs{x}_{i}\right\} )$ can be expressed in terms of the expectation values of higher products of the momentum occupation number operators 
\be
\left \langle\prod_{j=1}^{n}(n(\bs{k}_{j})+n(-\bs{k}_{j})+1)\right \rangle_{0}, 
\ee
which in turn can be written as functions of the initial correlation functions 
\be
\left \langle\prod_{j=1}^{n}\tilde{\phi}_{\bs{k}_{j}}^{(\sigma_{j})}\tilde{\phi}_{-\bs{k}_{j}}^{(\sigma_{j})}\right \rangle_{0}, 
\ee
where $\sigma_{j}=0,1$ and we denote $\tilde{\phi}_{\bs{k}_{j}}^{(0)}\equiv\tilde{\phi}_{\bs{k}_{j}}$,
$\tilde{\phi}_{\bs{k}_{j}}^{(1)}\equiv\tilde{\pi}_{\bs{k}_{j}}=\dot{\tilde{\phi}}_{\bs{k}_{j}}$. 
The result can be written in coordinate space as a weighted spatial average of the correlation functions 
\be
\left\langle\prod_{j=1}^{n}\tilde{\phi}^{(\sigma_{j})}(\bs{s}_{j}+\bs{x}_{2j-1})\tilde{\phi}^{(\sigma_{j})}(\bs{s}_{j}+\bs{x}_{2j})\right\rangle_{0}, 
\ee
which in the thermodynamic limit $L\to\infty$, due to the cluster decomposition property of the initial state, tends to the fully disconnected expression
\be
\prod_{j=1}^{n}\langle\tilde{\phi}(\bs{x}_{2j-1})\tilde{\phi}(\bs{x}_{2j})\rangle_{0},
\ee 
that is the gaussian result based only on the two-point correlation functions, precisely equal to the GGE prediction.

\section{Massless non-interacting post-quench hamiltonian}\label{sec:massless}

We now discuss the case of massless free evolution. This case should be studied separately, because the function $H(\bs{r})$ does not decay exponentially with the distance as before, and this property is important in the calculation of the $L$-scaling of spatial averages of initial correlation functions (Appendix~\ref{app:4}). Instead of decaying exponentially, $H(\bs{r})$ exhibits infrared divergences in $d \leq 2$, meaning that it diverges with $L$, while in $d>2$ it decays as a power of the distance
\be
H(\bs{r}) \sim \frac1{ |\bs{r}|^{d-2}} \quad \text{ for } |\bs{r}|\to\infty \text{ and } d>2.
\ee
More specifically, its precise functional form in dimensions $d=1,2$ and 3 is the following
\be
H(\bs{r}) = H(r) = 
\begin{cases}
({L}/{\pi}) - ({r}/{2})  & \text{ if } d=1, \\
\log(L/r) /(2\pi)  & \text{ if } d=2, \\
1/( 4\pi r) & \text{ if } d=3.
\end{cases}
\ee
This has as a direct consequence that the scaling behaviour of $\bar{C}^{(2)}(\bs{x}_{1},\bs{x}_{2})$ and $\bar{C}^{(4)}(\bs{x}_{1},\bs{x}_{2},\bs{x}_{3},\bs{x}_{4})$ is also different, depending on the dimensionality. Below we discuss separately the cases of $d=1,2$ and 3.

\subsection{One dimension}

In $d=1$, neither the two-point function nor the four-point function of the bosonic field $\phi$ become stationary in the large time limit, independently of the behaviour of the initial correlation functions $C_{0}^{(2)}$, $\ddot{C}_{0}^{(2)}$, $C_{0}^{(4)}$, $\ddot{C}_{0}^{(4)}$ and $\ddddot{C}_{0}^{(4)}$, since both $\bar{C}^{(2)}(\bs{x}_{1},\bs{x}_{2})$ and $\bar{C}^{(4)}(\bs{x}_{1},\bs{x}_{2},\bs{x}_{3},\bs{x}_{4})$ diverge in the thermodynamic limit $L\to\infty$. This observation is in agreement with what is known from \cite{cc06} for initial states that belong to the class of conformally invariant boundary states (which are gaussian), where it is shown that the two-point function $C_{t}^{(2)}(\bs{x}_{1},\bs{x}_{2})$ increases linearly with time. The divergence of $\bar{C}^{(2)}(\bs{x}_{1},\bs{x}_{2})$ with $L$, following the arguments in Section~\ref{sec:eq+GGE4}, reflects exactly this fact. However the $d=1$ case is special in the sense that local physical observables in a massless one-dimensional field theory, i.e. in a Conformal Field Theory, are described by correlations of \emph{vertex operators} instead of the free bosonic field $\phi$ and those correlation functions \emph{do} equilibrate (to values given by an effective thermal ensemble \cite{cc06} in some special cases of initial states, or to the GGE in more general cases \cite{cardy12,GGE-CFT}).

\subsection{Two dimensions}

In dimensions $d\geq 2$ the $\phi$-field correlations are physical observables themselves. 
In $d=2$, due to the logarithmic divergence of $H(r)$ with $L$, long time averaged correlation functions also diverge with $L$. Therefore, according to the above, there is no equilibration but instead correlations increase with time. Assuming that the initial correlation functions $C_{0}^{(2)}$ and $\ddot{C}_{0}^{(2)}$ do \emph{not} diverge with $L$ themselves, which is always true for ground states of non-critical pre-quench hamiltonians, the leading order values of long time averages are dominated by the contribution of terms with the most $H$-weighted spatial integrals, i.e. 
\begin{align}
\bar{C}^{(2)}(\bs{r}) &    \overset{L\to \infty}{\longrightarrow}
\frac{1}{2} \int\limits_{L^d} d^d s\, H(\bs{s}) \ddot{C}_{0}^{(2)}(\bs{r}-\bs{s}),
\end{align}
and
\begin{align}
& \bar{C}^{(4)}(\bs{x}_{1},\bs{x}_{2},\bs{x}_{3},\bs{x}_{4})  \nonumber \\
 & = \frac{1}{32 L^{2d}}\sum_{{\text{all perm.s}\atop \text{of }1,2,3,4}}  \iint\limits_{L^d} d^ds d^dr d^ds' d^dr' \,  H(\bs{s}')H(\bs{r}')  \nonumber \\
 & \times \ddddot{C}_{0}^{(4)}(\bs{s},\bs{s}'+\bs{s}+\bs{x}_{1}-\bs{x}_{2},\bs{r},\bs{r}'+\bs{r}+\bs{x}_{3}-\bs{x}_{4}) ,\label{eq:M2dD}
\end{align}
which, using the same arguments 
based on the cluster decomposition principle (Appendix~\ref{app:4}), reduces to
\begin{align}
& \bar{C}^{(4)}(\bs{x}_{1},\bs{x}_{2},\bs{x}_{3},\bs{x}_{4})  \nonumber \\
 &  \overset{L\to \infty}{\longrightarrow} \frac{1}{32} \sum_{{\text{all perm.s}\atop \text{of }1,2,3,4}}  \int\limits_{L^d}d^ds'H(\bs{s}')\ddot{C}_{0}^{(2)}(\bs{x}_{1}-\bs{x}_{2}+\bs{s}')  \nonumber \\
 & \times\int\limits_{L^d} d^dr'H(\bs{r}')\ddot{C}_{0}^{(2)}(\bs{x}_{3}-\bs{x}_{4}+\bs{r}').
\label{eq:M2dD}
\end{align}
If, on the other hand, $C_{0}^{(2)}$ also diverges with $L$, which may happen when the pre-quench hamiltonian is critical too, and if it diverges also logarithmically, while the other initial correlation functions do not diverge with $L$, then the contribution of the terms involving $C_{0}^{(2)}$ may be of equal importance with the rest, in which case we obtain expressions similar to (\ref{eq:Cinfty2}) and (\ref{eq:C4a}), i.e. those holding in the massive-to-massive case. Similar observations of divergence of the two-point function in the special case of free-to-free quantum quenches in dimensions $d=2$ have been made in Ref.~\cite{SC}, where it was shown that a mass quench in the Klein-Gordon theory leads to a two-point function that increases logarithmically with time.

\subsection{Three dimensions}

The $d=3$ case is different from the above, because $H(r)$ decays with the distance as a power and therefore the answer to the question whether the correlation functions equilibrate or not depends also on the large distance decay of the initial correlations and in particular whether they decay sufficiently fast so that all spatial integrals are convergent. For the two-point function, if $\ddot{C}_{0}^{(2)}$ decays exponentially, as is the case of a non-critical pre-quench hamiltonian, then $C_{t}^{(2)}$ equilibrates. The same is true if $\ddot{C}_{0}^{(2)}$ decays as a power but sufficiently fast, more specifically faster than $r^{-2}$, while if it decays as $r^{-2}$ or slower then $C_{t}^{(2)}$ does not equilibrate but increases with $t$ (i.e. $\bar{C}^{(2)}$ diverges with $L$). Similarly for the four-point function $C_{t}^{(4)}$, the large distance decay of $\ddot{C}_{0}^{(2)}$ and $\ddddot{C}_{0,\text{c}}^{(4)}$ determines if $C_{t}^{(4)}$ equilibrates or not. Notice that, especially in the critical pre-quench case, it is possible that some correlation functions equilibrate while others do not.

\subsection{General remarks for $d\geq 2$}

\begin{figure}[h!bt]
\begin{center}
\includegraphics[width=.4\textwidth]{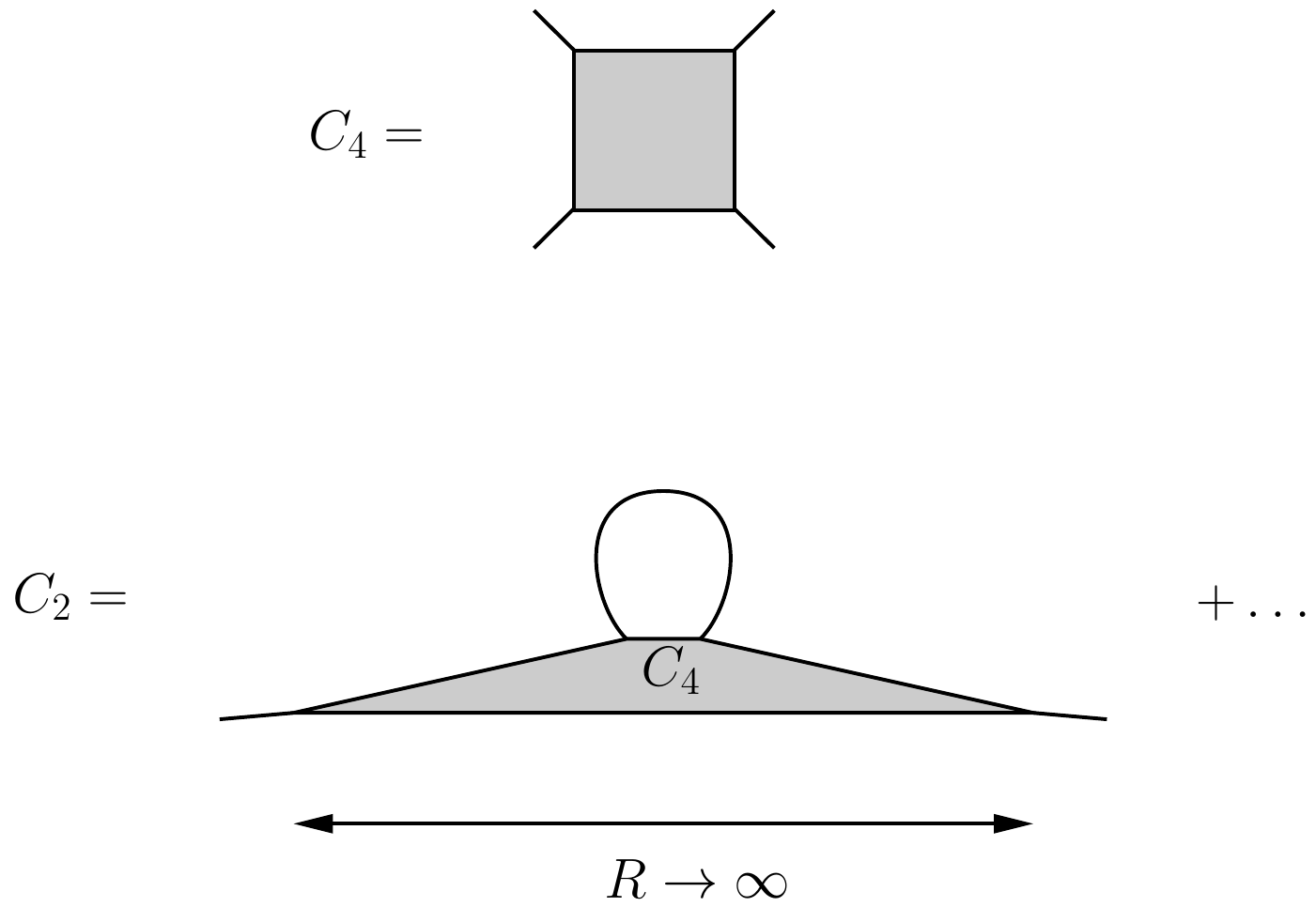}
\caption{Pictorial representation of the argument that, if the connected four-point function decays as a power when any of the coordinates goes to infinity, then the same holds also for the two-point function.}
\label{fig:FD1}
\end{center}
\end{figure}

According to the above, the scaling of long time averages of correlation functions as $L\to\infty$ in dimensions $d\geq 2$ (and therefore also the scaling of correlations as $t\to\infty$) depends highly on the large distance behaviour of the initial correlation functions $C_{0}^{(2)}$, $\ddot{C}_{0}^{(2)}$, $C_{0,c}^{(4)}$, $\ddot{C}_{0,c}^{(4)}$ and $\ddddot{C}_{0,c}^{(4)}$. This is determined by the Renormalisation Group (RG) analysis of the pre-quench hamiltonian. Note that the large distance behaviour of initial $\pi$-field correlations is related to that of the $\phi$-field correlations, since $\pi=\dot\phi$. In the case of the two-point function, for example, $\pi$-field correlations can be expressed, using general properties like space-time symmetries before the quench and Lorentz invariance, as a spatial derivative of the $\phi$-field correlations
\be
\ddot C_0^{(2)}(r) = - \frac1{r} \frac{\partial^2}{\partial r^2} C_0^{(2)}(r) .
\ee
Also note that the large distance decay of initial connected correlation functions of different orders is not independent of one another. If the four-point function ${C}_{0,\text{c}}^{(4)}$ decays as a power, the same must hold also for the two-point function ${C}_{0}^{(2)}$. Indeed, one contribution to the diagrammatic expansion of the two-point function involves the four-point function with two legs contracted forming a loop. This diagram decays as a power of the distance of the two uncontracted points and since the large distance decay is determined by the slowest decaying contribution, the full two-point function must decay as a power (Fig.~\ref{fig:FD1}). It should be mentioned that this argument is not perturbative. 
In general we can distinguish two typical types of behaviour: initial connected correlation functions decay exponentially with the distance (as for non-critical pre-quench hamiltonians) or they decay algebraically (as for critical pre-quench hamiltonians). In the critical case it is also possible that correlation functions diverge with $L$ themselves. 

In all of the above studied cases, both in $d=2$ and $d=3$, whenever all correlation functions equilibrate, their stationary values satisfy Wick's theorem and are given by the GGE. Moreover, in the non-critical pre-quench case, even when correlation functions do not equilibrate, they are still expressed in terms of \emph{solely} the two-point initial correlation functions ($C_{0}^{(2)}$ and $\ddot{C}_{0}^{(2)}$) \emph{at leading order} in the large time and thermodynamic limit. Accordingly, the GGE predicts correctly (only) the leading order of long time averaged correlation functions in this case. Higher order correlation functions ($C_{0,c}^{(4)}$, $\ddot{C}_{0,c}^{(4)}$ and $\ddddot{C}_{0,c}^{(4)}$) give contributions next-to-leading order at large $L$, which may still diverge with $L$ but slower. Therefore, even though Wick's theorem (\ref{eq:C4ba}) is \emph{not} valid in this case, it is still applicable in the following weak sense:
\begin{align}
\frac{\bar{C}^{(4)}(\bs{x}_{1},\bs{x}_{2},\bs{x}_{3},\bs{x}_{4})}{\bar{C}^{(2)}(\bs{x}_{1},\bs{x}_{2}) \bar{C}^{(2)}(\bs{x}_{3},\bs{x}_{4})+[2\leftrightarrow3]+[2\leftrightarrow4]}  \overset{L\to \infty}{\longrightarrow} 1,
 \label{eq:Wick}
\end{align}
Note that the difference between numerator and denominator may be divergent with $L$. Following the arguments of Section~\ref{sec:hcf} and Appendix~\ref{app:4}, this weak version of Wick's theorem is expected to apply similarly to all higher order correlation functions. The above result is always true when the pre-quench hamiltonian is non-critical, but it is also true in the case of critical pre-quench hamiltonians for a wide range of power law exponents for the correlation functions. Even though a general classification of all possibilities is beyond the objectives of the present work, possible exceptions may arise only in marginal cases when the exponents of the algebraic decay of different correlation functions have special values, such that the contributions of connected four-point correlations functions survive at leading order in the large time and thermodynamic limit. 

It should also be emphasised that for any specific model, as the dimensionality $d$ increases the deviations from the GGE become smaller and smaller until they finally vanish above some critical dimension. This is due to the scaling of the convolutions involved in the expressions for the long time averaged correlation functions: both the function $H(r)\sim 1/r^{d-2}$ and the initial correlation functions decay faster as $d$ increases (even if they decay as a power law), so that the expressions above eventually become convergent. This can be seen easier in momentum space where the $L\to\infty$ divergences of the convolutions are determined by the order of the pole at $k=0$. While the poles of $\tilde{H}(\bs{k})$ and of the correlation functions are fixed, the phase space integration measure $d^dk \sim k^{d-1} dk$ changes with $d$ and eventually cancels the pole at $k=0$. Therefore \emph{equilibration occurs and the GGE conjecture becomes correct for any interacting-to-free quench above some critical dimension}. This is obviously the same argument as in mean field theory, which (both in the dynamical and in the equilibrium version) becomes exact above the critical dimension.

\section{Conclusions \& Outlook}
\label{sec:c+o}

In this work we have studied the equilibration properties and validity of the GGE conjecture in higher dimensional bosonic QFTs, undergoing a quantum quench from a free or interacting hamiltonian to a free hamiltonian. If the post-quench mass is nonzero the system equilibrates, its correlation functions become gaussian i.e. satisfy Wick's theorem and the GGE predicts correctly their stationary values. We can interpret this effect as loss of memory of initial non-gaussian correlations, due to evolution under a gaussian massive hamiltonian. From the technical point of view, considering the combination of thermodynamic and large time limit, it turns out that only partial information about the initial state is required to describe local observables and this information is precisely the initial two-point function averaged over space with a weight function $H(r)$ dependent on the post-quench hamiltonian. If the post-quench hamiltonian is massless, despite the fact that the evolution may lead to unbounded increase of correlation functions, the GGE conjecture still gives correct predictions for the leading order correlations in the thermodynamic and large time limit, at least (but not only) in the case of non-critical pre-quench hamiltonians. Contributions of non-gaussian initial correlations may still increase with time or system size but slower, so that their strength relative to the leading order decays to zero. Moreover, above some model-dependent critical dimension, corrections to the GGE predictions vanish and the GGE conjecture becomes exact. Notice that our arguments are not sensitive to the particular quench under consideration, but refers to a large class of quenches and is expected to be valid for non-relativistic models as well.

It is tempting to attempt to generalise our approach to the case of interacting evolution in order to test the hypothesis that in dimensions $d>1$ a suitable GGE, built with approximately conserved charges, describes the pre-thermalisation regime. This corresponds to the intermediate stage of the time evolution, after dephasing of quasi-particles takes place and before thermalisation due to their interaction is finally reached \cite{prethrm1,prethrm2,prethrm3,prethrm4,prethrm5,prethrm6,ekmr-14,mgs,prethrm9,prethrm10}. Our method relies on two key features of free evolution that are not generally valid in the interacting case: the solution of the equations of motion for the field operators and the locality of these fields, on which the cluster decomposition is based. However, perturbation theory suggests that these properties may be unaffected by the introduction of weak interactions, as long as: the dimensionality is $d>1$, we are away from criticality and excitations have a finite maximum group velocity. These conditions are expected to guarantee the light-cone form of dynamics and the possibility to express the correlation functions in terms of local initial fields, which are sufficient conditions for the application of our method. In this perturbative approach, the approximate conserved charges would not be quadratic operators in terms of the local fields, which means that they would not be necessarily local either. We hope to come back to this problem in a future publication.

Interesting questions arise also in the context of supersymmetric quantum field theories in $D=d+1>2$ which are higher dimensional yet integrable. The natural candidates to explore the problem of equilibration in such theories are the ${\cal N}=4$ Super-Yang-Mills (SYM) and ${\cal N}=6$ Super-Chern-Simons theories (ABJM theory), which are conformal and in some sense also integrable \cite{Beisert:2010jr}, though integrability in this case means partial knowledge of the spectrum of the theories, but not existence of infinite number of \emph{local} conserved charges, as in one dimension. Indeed an infinite tower of \emph{non-local} conserved charges appears in the scattering amplitudes of gluons (Yangian Symmetry \cite{Drummond:2009fd}), but the role of these charges in determining the spectrum is not clear. There are two alternative approaches to study the quench dynamics in such models. One is based on a mapping of the quench problem to a boundary problem in a Euclidean slab geometry \cite{cc06}. This was initially applied to $1d$ Conformal Field Theory with a conformally invariant initial state which turns out to lead to effective thermalisation, while it was later generalised to perturbations of the latter states \cite{cardy12,GGE-CFT} in which case a suitable GGE was shown to be correct. A natural extension to supersymmetric models would be to consider supersymmetric initial states \cite{Gaiotto:2008sa} where powerful tools such as localisation \cite{Pestun:2007rz} are available. Another approach studied extensively is based on the AdS/CFT correspondence \cite{qe-hol,qe-hol1,qe-hol2,qe-hol3,qe-hol4,qe-hol5,qe-hol6,qe-hol7,myers} where equilibration is thermal. An interesting open question is: if a thermal ensemble is described in this context by a black hole, what is the gravity-dual of a GGE ensemble? It is conjectured \cite{G1,G2,GGE-CFT} that it corresponds to a higher spin black hole, whereas in classical gravity black holes can only have a small number of additional conserved charges (mass, angular momentum and electric charge) in contrast to the infinite number associated with integrability. It would be interesting to use our approach in order to study such questions and compare with the other available methods.

\section*{Acknowledgments} We are grateful to Pasquale Calabrese, John Cardy, Mario Collura, Andrea Gambassi, Matteo Marcuzzi and Domenico Seminara for helpful discussions. 
This work was supported by the ERC under Starting Grant 279391 EDEQS (both SS and GM).



\appendix

\section{Preliminary definitions}
\label{app:-1}

We consider the Klein-Gordon field described by the hamiltonian (\ref{hamiltonian}). The time-evolved field operator $\phi$ in the Heisenberg picture is
\begin{align}
& \phi(\bs{x};t) = \frac{1}{L^{d/2}}\sum_{\bs{k}}{e^{i\bs{k}\cdot\bs{x}}\tilde{\phi}_{\bs{k}}(t)} \nonumber \\
& =\frac{1}{L^{d/2}}\sum_{\bs{k}}{e^{i\bs{k}\cdot\bs{x}}\;\frac{1}{\sqrt{2E_{\bs{k}}}}(a_{\bs{k}}e^{-iE_{\bs{k}}t}+a_{-\bs{k}}^{\dagger}e^{+iE_{\bs{k}}t})},
\end{align}
and similarly the conjugate momentum $\pi$ is
\begin{align}
& \pi(\bs{x};t)=\frac{1}{L^{d/2}}\sum_{\bs{k}}e^{i\bs{k}\cdot\bs{x}}\tilde{\pi}_{\bs{k}}(t) \nonumber \\
& =\frac{(-i)}{L^{d/2}}\sum_{\bs{k}}e^{i\bs{k}\cdot\bs{x}}\; \sqrt{\frac{E_{\bs{k}}}{2}}(a_{\bs{k}}e^{-iE_{\bs{k}}t}-a_{-\bs{k}}^{\dagger}e^{+iE_{\bs{k}}t}).
\end{align}
Here $L$ is the system size and we assume periodic boundary conditions, so that the momenta are given by $\bs{k}=2\pi n/L$ with $n$ integer. 

The creation-annihilation operators are
\begin{align}
a_{\bs{k}} & =\sqrt{\frac{E_{\bs{k}}}{2}}\tilde{\phi}_{\bs{k}}(0)+\frac{i}{\sqrt{2E_{\bs{k}}}}\tilde{\pi}_{\bs{k}}(0), \nonumber \\
a_{-\bs{k}}^{\dagger} & =\sqrt{\frac{E_{\bs{k}}}{2}}\tilde{\phi}_{\bs{k}}(0)-\frac{i}{\sqrt{2E_{\bs{k}}}}\tilde{\pi}_{\bs{k}}(0).
\label{app:eq:3}
\end{align}

The field $\phi(\bs{x};t) $ automatically satisfies the equations of motion, which in momentum space read
\be
\ddot{\tilde{\phi}}_{\bs{k}}(t) + E_{\bs{k}}^2 \tilde{\phi}_{\bs{k}}(t) = 0,
\ee
with the obvious solution
\be
\tilde{\phi}_{\bs{k}}(t) = \tilde{\phi}_{\bs{k}}(0) \cos E_{\bs{k}} t + \tilde{\pi}_{\bs{k}}(0) \frac{\sin E_{\bs{k}} t }{E_{\bs{k}}}. \label{app:sol}
\ee

Before we proceed further, we also report an identity that will be useful later in expressing the time averaged values of field correlations in terms of their initial values in coordinate space. From (\ref{app:eq:3}) the operator $n(\bs{k})+n(-\bs{k})$ can be written in terms of the field $\phi$ and its time derivative $\dot{\phi}=\pi$ as
\begin{align}
  & n(\bs{k})+n(-\bs{k})=a_{\bs{k}}^{\dagger}a_{\bs{k}}+a_{-\bs{k}}^{\dagger}a_{-\bs{k}}
 \nonumber \\
 & =E_{\bs{k}}\tilde{\phi}_{\bs{k}}\tilde{\phi}_{-\bs{k}}+\tilde{\pi}_{\bs{k}}\tilde{\pi}_{-\bs{k}}/E_{\bs{k}}-1,
 \label{app:eq:1}
\end{align}
where in the last step we used the commutation relations $[\tilde{\phi}_{\bs{k}},\tilde{\pi}_{\bs{q}}]=i\delta_{\bs{k},-\bs{q}}$
and $[\tilde{\phi}_{\bs{k}},\tilde{\phi}_{\bs{q}}]=[\tilde{\pi}_{\bs{k}},\tilde{\pi}_{\bs{q}}]=0$.

\section{The one-point correlation function}
\label{app:0}

Using the above expressions for the evolved fields (Appendix~\ref{app:-1}), we can easily find the evolution of the one-point function. This is only non-zero if its initial value $ \phi_0\equiv\langle\phi(\bs{x};0)\rangle $ is non-zero (as for example in the case of the ground state of a QFT with broken symmetry). Indeed, from (\ref{app:sol}) we obtain 
\begin{align}
 & C_{t}^{(1)}(\bs{x})\equiv \langle\phi(\bs{x};t)\rangle 
 = \frac{1}{L^{d/2}} \sum_{\bs{k}}  {  e^{i\bs{k}\cdot\bs{x}} \; \langle \tilde{\phi}_{\bs{k}}(0) \rangle  \cos E_{\bs{k}} t } \nonumber \\
 & = \phi_0 \cos m t,
 \label{ap:C1}
\end{align}
since 
\be
\langle\tilde{\phi}_{\bs{k}}(0)\rangle = \frac{1}{L^{d/2}} \int_{L^d}{d^dx \; e^{-i\bs{k}\cdot\bs{x}} \langle\phi(\bs{x};0)\rangle = \phi_0 \; L^{d/2}\delta^{(d)}_{\bs{k},\bs{0}} }  ,
\ee
and $ \langle \tilde{\pi}_{\bs{k}}(0) \rangle = ({d}/{dt})\langle \tilde{\phi}_{\bs{k}}(t) \rangle |_{t=0}= 0 $ as a consequence of the continuity of the time derivative of the field at $t=0$ and the fact that before the quench there was no time evolution, since the initial state was an eigenstate of the pre-quench hamiltonian. 
From (\ref{ap:C1}) we see that the evolution of the one-point function is trivially oscillating between the values $\pm \phi_0$. 

In the above and in what follows we make the assumption that the initial state is translationally invariant. 
Without loss of generality, from now on we will also assume that the initial value of the field is zero, i.e. $\phi_0=0$. In the opposite case, we can always redefine the field $\phi$ to be the difference from the above one-point function, therefore focusing on the quantum fluctuations about this value.

\section{The two-point correlation function}
\label{app:1}

Next we calculate the two-point function at equal times in terms of expectation values in the initial state
\begin{align}
 & C_{t}^{(2)}(\bs{x},\bs{y})\equiv\langle\phi(\bs{x};t)\phi(\bs{y};t)\rangle \nonumber \\ 
 &=\frac{1}{L^d}\sum_{\bs{k}_{1},\bs{k}_{2}}\frac{1}{\sqrt{2E_{\bs{k}_{1}}}}\frac{1}{\sqrt{2E_{\bs{k}_{2}}}}\; e^{i\bs{k}_{1}\cdot\bs{x}+i\bs{k}_{2}\cdot\bs{y}} \nonumber \\ 
 & \times\Big[\langle a_{\bs{k}_{1}}a_{\bs{k}_{2}}\rangle_{0}\; e^{-i(E_{\bs{k}_{1}}+E_{\bs{k}_{2}})t}+\langle a_{-\bs{k}_{1}}^{\dagger}a_{\bs{k}_{2}}\rangle_{0}\; e^{+i(E_{\bs{k}_{1}}-E_{\bs{k}_{2}})t} \nonumber \\ 
&  +\langle a_{\bs{k}_{1}}a_{-\bs{k}_{2}}^{\dagger}\rangle_{0}\; e^{-i(E_{\bs{k}_{1}}-E_{\bs{k}_{2}})t}+\langle a_{-\bs{k}_{1}}^{\dagger}a_{-\bs{k}_{2}}^{\dagger}\rangle_{0}\; e^{+i(E_{\bs{k}_{1}}+E_{\bs{k}_{2}})t}\Big],\nonumber
\end{align}
where the index zero means that the expectation values are calculated
in the initial state. Using our assumption that the initial state is translationally invariant, 
the above expectation values are nonzero only when the momenta 
$\bs{k}_{1},\bs{k}_{2}$ are equal or opposite. We therefore have 
\begin{align}
& C_{t}^{(2)}(\bs{x},\bs{y}) =\frac{1}{L^d}\sum_{\bs{k}}\frac{1}{2E_{\bs{k}}}\; e^{i\bs{k}\cdot(\bs{x}-\bs{y})} \nonumber \\ 
& \times \Big[1+F(-\bs{k})+F(\bs{k})+G(-\bs{k})\; e^{-2iE_{\bs{k}}t}+G^*(-\bs{k})\; e^{+2iE_{\bs{k}}t}\Big].
\end{align}
where $G(\bs{k})$ and $F(\bs{k})$ are functions that depend on the particular
initial state: $F(\bs{k})$ expresses the momentum distribution in the initial state
\begin{equation}
F(\bs{k})=\langle a_{\bs{k}}^{\dagger}a_{\bs{k}}\rangle_{0}=\langle n(\bs{k})\rangle_{0},
\end{equation}
and it is a real function, while $G(\bs{k})$ gives the probability amplitude for the presence of a pair of opposite momentum particles in the initial state
\begin{equation}
G(\bs{k})=\langle a_{-\bs{k}}a_{\bs{k}}\rangle_{0},
\end{equation}
and it is an even function. We will also assume that the initial state is rotationally invariant, in which case the above are functions of the norm $k=|\bs{k}|$ but not of the direction of the momenta $\bs{k}$. 

In the thermodynamic limit $L\to\infty$, the sum in the above expression
becomes an integral. Using the translational and rotational symmetry of the initial state we can further simplify the above expressions, since the functions $F$ and $G$ are function of $|\bs{k}|$ only. 
We now take the large time limit $t\to\infty$ of the correlation function.
In the case of massive post-quench dispersion relation, $E_{k}=\sqrt{k^{2}+m^{2}}$
with $m\neq0$, the stationary phase method shows that the oscillating
time dependent terms vanish and therefore the two-point
function equilibrates. In the massless case the large time asymptotics of the above correlation function is sensitive to the behaviour of the function $G(k)$. Typically, if $G(k)$ decays sufficiently fast at large $k$ and does not have a pole at $k=0$ but possibly other singularities elsewhere in the complex $k$-plane, then the time dependent terms decay for large times and the correlation function equilibrates.

As long as the two-point function equilibrates, its stationary value is equal to its long time average value which is
\begin{align}
\bar{C}^{(2)}(\bs{r}) & = \frac{1}{L^d}\sum_{\bs{k}} \frac{1}{2E_{k}}e^{i\bs{k}\cdot\bs{r}}{(1+2F(k))}.
\label{app:eq:2pt_lt}
\end{align}
Note that it depends solely on the initial expectation values of the occupation numbers of the momentum modes $F(k)=\langle n(\bs{k})\rangle_{0}$.


Using the identity (\ref{app:eq:1}) in (\ref{app:eq:2pt_lt}) we can rewrite it as
\begin{align}
& \bar{C}^{(2)}(\bs{r}) =\frac{1}{2}\bigg(C_{0}^{(2)}(\bs{r}) +\int\limits_{L^d} d^ds\, H(\bs{s})\ddot{C}_{0}^{(2)}(\bs{r} -\bs{s})\bigg),
\end{align}
where the functions $H$ and $\ddot{C_{0}}^{(2)}$ have been defined in (\ref{eq:Ha}) and (\ref{eq:ddC})
\begin{align}
H(\bs{x})\equiv & \int\frac{d^dk}{(2\pi)^d}\frac{e^{i\bs{k}\cdot\bs{x}}}{E_{k}^{2}},\label{app:eq:H}\\
C_{0}^{(2)}(\bs{x})\equiv & C_{0}^{(2)}(\bs{0},\bs{x}),\label{app:eq:C}
\end{align}
and 
\begin{align}
& \ddot{C}_{0}^{(2)}(\bs{x}) \equiv\left\langle \pi(\bs{0},0)\pi(\bs{x},0)\right\rangle \nonumber \\
& =\left.\frac{\partial}{\partial t_{1}}\frac{\partial}{\partial t_{2}}\left\langle \phi(\bs{0},t_{1})\phi(\bs{x},t_{2})\right\rangle \right|_{t_{1}=t_{2}=0}.
\end{align}

\section{The four-point correlation function}
\label{app:2}

We proceed to the calculation of the four-point function. The equal time four-point function can be expressed as
\begin{align}
& C_{t}^{(4)}(\bs{x}_{1},\bs{x}_{2},\bs{x}_{3},\bs{x}_{4}) \equiv\langle\phi(\bs{x}_{1};t)\phi(\bs{x}_{2};t)\phi(\bs{x}_{3};t)\phi(\bs{x}_{4};t)\rangle \nonumber \\
& = \frac{1}{L^{2d}}\sum_{\bs{k}_{1},\bs{k}_{2},\bs{k}_{3},\bs{k}_{4}}\frac{1}{4\sqrt{\prod_{i=1}^{4}E_{\bs{k}_{i}}}}\; e^{i\sum_{i=1}^{4}\bs{k}_{i}\cdot\bs{x}_{i}}  \nonumber \\
& \times \sum_{{\text{all }\{\sigma_{i}\}\atop \sigma_{i}=\pm}}\left\langle \prod_{i=1}^{4}a_{-\sigma_{i}\bs{k}_{i}}^{(\sigma_{i})}\right\rangle _{0}\; e^{i\sum_{i=1}^{4}\sigma_{i}E_{\bs{k}_{i}}t},\label{app:4pt_1}
\end{align}
where we used the compact notation $a_{\bs{k}}^{(+)}\equiv a_{\bs{k}}^{\dagger}$
and $a_{\bs{k}}^{(-)}\equiv a_{\bs{k}}$. 

As before, assuming that the initial expectation values are suitably smooth functions so that the oscillating terms decay for large times, the four-point function equilibrates and its stationary value is given by the terms that are not accompanied by time-oscillating phases, i.e. those satisfying the condition 
 $\sum_{i=1}^{4}\sigma_{i}E_{\bs{k}_{i}}=0$ 
with $\sigma_{i}=\pm$. If the four-point function does not equilibrate, then this expression gives the long time average. Moreover, by translational invariance of the initial state, the above expectation values are non zero only if their total momentum is zero 
 $\sum_{i=1}^{4} \bs{k}_{i}=0$. 
Taking into account that the momenta $\bs{k}_{i}$ are quantised, i.e. each of their components is an integer multiple of $2\pi/L$ and that the first equation is not a linear equation (since $E_{\bs{k}} = \sqrt{|\bs{k}|^{2}+m^{2}} $), the terms that give non-vanishing contributions in the thermodynamic limit are those with
equal number of $a$ and $a^{\dagger}$ operators, i.e. those for which $\sum_{i=1}^{4}\sigma_{i}=0$ (which are $4!/(2! 2!)=6$ in number, out of $2^{4}=16$) and for which the 
momenta of the $a$ operators match with those of the $a^{\dagger}$
operators in pairs of opposite values. All other solutions in the set of integer numbers are sparse and do not contribute in the thermodynamic limit. 

Considering all possible permutations we finally obtain an expression in terms of the initial expectation values of the occupation numbers $n(\bs{k})$ and their products
\begin{align}
&  \bar{C}^{(4)}(\bs{x}_{1},\bs{x}_{2},\bs{x}_{3},\bs{x}_{4})  \nonumber \\
& = \frac{1}{2}\frac{1}{L^{2d}}\sum_{\bs{k},\bs{p}}\frac{1}{4E_{\bs{k}}E_{\bs{p}}}
S(\bs{k},\bs{p};\bs{x}_{1},\bs{x}_{2},\bs{x}_{3},\bs{x}_{4}) \nonumber \\
& \times  \left(\langle n(\bs{k})n(\bs{p})\rangle_{0}+\langle n(\bs{k})\rangle_{0}+\frac{1}{4}\right),
 \label{app:eq:lt1}
\end{align}
where the function $S(\bs{k},\bs{p};\bs{x}_{1},\bs{x}_{2},\bs{x}_{3},\bs{x}_{4})$ is defined in (\ref{S-function})
\be
S(\bs{k},\bs{p};\bs{x}_{1},\bs{x}_{2},\bs{x}_{3},\bs{x}_{4})\equiv\sum_{{\text{all perm.s}\atop \text{of 1,2,3,4}}}e^{i\bs{k}\cdot(\bs{x}_{2}-\bs{x}_{1})+i\bs{p}\cdot(\bs{x}_{4}-\bs{x}_{3})}.
\ee
In the thermodynamic limit we replace ${L^{-2d}}\sum_{\bs{k},\bs{p}} \to {(2\pi)^{-2d}}\int{d^d{k}\, d^d{p}} $.

Finally, we write the last expression in terms of coordinate space initial correlation functions. The function $S(\bs{k},\bs{p};\bs{x}_{1},\bs{x}_{2},\bs{x}_{3},\bs{x}_{4})$ is even in both $\bs{k}$ and $\bs{p}$ and symmetric under their interchange. Therefore we can replace
the expression $\left(\langle n(\bs{k})n(\bs{p})\rangle_{0}+\langle n(\bs{k})\rangle_{0}+\frac{1}{4}\right)$
in the sum by 
\begin{align}
 & \frac{1}{4}\left\langle \left(n(\bs{k})+n(-\bs{k})+1\right)\left(n(\bs{p})+n(-\bs{p})+1\right)\right\rangle _{0}=\nonumber\\
 & =\frac{1}{4}\big(\left\langle n(\bs{k})n(\bs{p})\right\rangle _{0}+\left\langle n(-\bs{k})n(\bs{p})\right\rangle _{0} + \left\langle n(\bs{k})n(-\bs{p})\right\rangle _{0}
 \nonumber\\
 & +\left\langle n(-\bs{k})n(-\bs{p})\right\rangle _{0}+\left\langle n(\bs{k})\right\rangle _{0}+\left\langle n(-\bs{k})\right\rangle _{0}+\left\langle n(\bs{p})\right\rangle _{0}=\nonumber\\
 &+\left\langle n(-\bs{p})\right\rangle _{0}+1\big),
\end{align}
and use the identity (\ref{app:eq:1}) to write (\ref{app:eq:lt1}) as 
\begin{align}
&  \bar{C}^{(4)}(\bs{x}_{1},\bs{x}_{2},\bs{x}_{3},\bs{x}_{4})=
 \frac{1}{32}\frac{1}{L^{2d}}\sum_{\bs{k},\bs{p}} S(\bs{k},\bs{p};\bs{x}_{1},\bs{x}_{2},\bs{x}_{3},\bs{x}_{4}) \nonumber \\
 & \times \bigg(\left\langle \tilde{\phi}_{\bs{k}}\tilde{\phi}_{-\bs{k}}\tilde{\phi}_{\bs{p}}\tilde{\phi}_{-\bs{p}}\right\rangle _{0}+\frac{1}{E_{\bs{k}}^{2}}\left\langle \tilde{\pi}_{\bs{k}}\tilde{\pi}_{-\bs{k}}\tilde{\phi}_{\bs{p}}\tilde{\phi}_{-\bs{p}}\right\rangle _{0} \nonumber \\
 & +\frac{1}{E_{\bs{p}}^{2}}\left\langle \tilde{\phi}_{\bs{k}}\tilde{\phi}_{-\bs{k}}\tilde{\pi}_{\bs{p}}\tilde{\pi}_{-\bs{p}}\right\rangle _{0}+\frac{1}{E_{\bs{k}}^{2}E_{\bs{p}}^{2}}\left\langle \tilde{\pi}_{\bs{k}}\tilde{\pi}_{-\bs{k}}\tilde{\pi}_{\bs{p}}\tilde{\pi}_{-\bs{p}}\right\rangle _{0}\bigg).
\end{align}
Notice that the large time value of the four-point correlation function
depends solely on four-point initial correlations, not on lower order
correlations of the initial state. 

The last step is to write the correlation functions in coordinate space, thus obtaining
\begin{align}
& \bar{C}^{(4)}(\bs{x}_{1},\bs{x}_{2},\bs{x}_{3},\bs{x}_{4}) = \nonumber \\
& \frac{1}{32}\sum_{{\text{all perm.s}\atop \text{of } 1,2,3,4}}  \iint\limits_{L^d} \frac{d^dsd^dr }{L^{2d}} \, 
\Bigg [ C_{0}^{(4)} \left(\bs{s},\bs{s}+\bs{x}_{1}-\bs{x}_{2},\bs{r},\bs{r}+\bs{x}_{3}-\bs{x}_{4}\right) \nonumber \\
& + \int\limits_{L^d}d^ds' \, H(\bs{s}') \ddot{C}_{0}^{(4)} (\bs{s},\bs{s}'+\bs{s}+\bs{x}_{1}-\bs{x}_{2},\bs{r},\bs{r}+\bs{x}_{3}-\bs{x}_{4}) \nonumber \\
& + \int\limits_{L^d}d^dr' \, H(\bs{r}') \big(\ddot{C}_{0}^{(4)}(\bs{s},\bs{s}+\bs{x}_{1}-\bs{x}_{2},\bs{r},\bs{r}'+\bs{r}+\bs{x}_{3}-\bs{x}_{4})\big)^{*}  \nonumber \\
& + \iint\limits_{L^d}d^ds'd^dr' \, H(\bs{s}')H(\bs{r}') 
\ddddot{C}_{0}^{(4)}(\bs{s},\bs{s}'+\bs{s}+\bs{x}_{1}-\bs{x}_{2},\bs{r},\bs{r}'+\bs{r}+\bs{x}_{3}-\bs{x}_{4})  \Bigg ], \label{app:eq:app6a}
\end{align}
where $\ddot{C}_{0}^{(4)}$ and $\ddddot{C}_{0}^{(4)}$ are defined in (\ref{ddC4}) and (\ref{ddddC4}) 
\be
\ddot{C}_{0}^{(4)}(\bs{x}_1,\bs{x}_2,\bs{x}_3,\bs{x}_4) = \left\langle \pi(\bs{x}_{1})\pi(\bs{x}_{2})\phi(\bs{x}_{3})\phi(\bs{x}_{4})\right\rangle _{0}
\ee
and 
\be
\ddddot{C}_{0}^{(4)}(\bs{x}_1,\bs{x}_2,\bs{x}_3,\bs{x}_4) = \left\langle \pi(\bs{x}_{1})\pi(\bs{x}_{2})\pi(\bs{x}_{3})\pi(\bs{x}_{4})\right\rangle _{0}
\ee
and the function $H(\bs{x})$ has been defined in (\ref{eq:Ha}).

\section{Spatial averages of initial correlation functions}
\label{app:4}

We will show, using the cluster expansion and the cluster decomposition principle, that the spatial weighted averages of initial correlation functions entering in (\ref{eq:6a}) are determined by the two-point functions only, provided that the weight function $H(r)$ decays exponentially with the distance $r$, or is, more generally, an integrable function in the whole infinite space. This argument is based on a multi-dimensional generalisation of the trivial fact that the infinite average value $\bar f$ of any function $f(x)$ equals its value at infinity, provided that it is finite, i.e.
\be
\bar{f} \equiv \lim_{L\to\infty} L^{-1} \int_0^L dx f(x) = f(+\infty).
\ee

Let us focus on the first term in the integral of (\ref{eq:6a}), which is essentially a spatial average of the initial four-point correlation function $C_{0}^{(4)}$ with respect to two coordinate variables $\bs{s}$ and $\bs{r}$. This is then equal to its value at $|\bs{s}|\to\infty$ and $|\bs{r}|\to\infty$, i.e. when the four points are separated in two pairs infinitely far from each other. According to the cluster decomposition principle, in this limit the correlation function tends to the disconnected expression, i.e. the product of two two-point correlation functions
\begin{align}
 & \lim_{L\to\infty} \iint\limits_{L^{d}} \frac{d^dsd^dr }{L^{2d}} \, C_{0}^{(4)}\left(\bs{s},\bs{s}+\bs{x}_{1}-\bs{x}_{2},\bs{r},\bs{r}+\bs{x}_{3}-\bs{x}_{4}\right) \nonumber \\
 & =  C_{0}^{(2)}(\bs{x}_{1}-\bs{x}_{2}) C_{0}^{(2)}(\bs{x}_{3}-\bs{x}_{4}).
\end{align}
In the above we used the fact that the initial expectation value of the field $\phi$ vanishes, $\langle\phi(\bs{x})\rangle_0=\phi_0=0$. 
Similar results hold for all other terms in the integral of (\ref{eq:6a}). For the last term, for example, we find 
\begin{align}
 & \lim_{L\to\infty} \iint\limits_{L^{d}} \frac{d^dsd^dr }{L^{2d}} \iint\limits_{L^d} d^ds'd^dr' \, H(\bs{s}')H(\bs{r}') \ddddot{C}_{0}^{(4)}(\bs{s},\bs{s}'+\bs{s}+\bs{x}_{1}-\bs{x}_{2},\bs{r},\bs{r}'+\bs{r}+\bs{x}_{3}-\bs{x}_{4}) \nonumber \\
  & = \iint d^ds' d^dr' \, H(\bs{s}') H(\bs{r}') \, \ddot{C}_{0}^{(2)}(\bs{x}_{1}-\bs{x}_{2}+\bs{s}') \ddot{C}_{0}^{(2)}(\bs{x}_{3}-\bs{x}_{4}+\bs{r}').
\end{align}
Note that the last expression is always finite for a massive post-quench hamiltonian, because the function $H(\bs{r})$ decays exponentially with the distance $r$.

More explicitly, the general cluster expansion for the $C_{0}^{(4)}$ correlation function is
\begin{align}
 & C_{0}^{(4)}\left(\bs{s}_1,\bs{s}_2,\bs{s}_3,\bs{s}_4\right) = C_{0}^{(2)}\left(\bs{s}_1,\bs{s}_2\right)C_{0}^{(2)}\left(\bs{s}_3,\bs{s}_4\right)  \nonumber \\
 & +C_{0}^{(2)}\left(\bs{s}_1,\bs{s}_3\right)C_{0}^{(2)}\left(\bs{s}_2,\bs{s}_4\right) + C_{0}^{(2)}\left(\bs{s}_1,\bs{s}_4\right)C_{0}^{(2)}\left(\bs{s}_2,\bs{s}_3\right) \nonumber \\
 & +C_{0,c}^{(4)}\left(\bs{s}_1,\bs{s}_2,\bs{s}_3,\bs{s}_4\right)
\end{align}
where $C_{0,c}^{(4)}$ is the connected initial four-point function. Applying this formula and taking into account the translational invariance of the initial state, we can write the integrand of the first term in (\ref{eq:6a}) as
\begin{align}
 & C_{0}^{(4)}\left(\bs{s},\bs{s}+\bs{x}_{1}-\bs{x}_{2},\bs{r},\bs{r}+\bs{x}_{3}-\bs{x}_{4}\right) \nonumber \\
 & =\underbrace{C_{0}^{(2)}(\bs{x}_{1}-\bs{x}_{2})C_{0}^{(2)}(\bs{x}_{3}-\bs{x}_{4})}_\text{type I} 
 \nonumber \\& +\underbrace{C_{0}^{(2)}(\bs{r}-\bs{s})C_{0}^{(2)}(\bs{r}-\bs{s}+\bs{x}_{3}-\bs{x}_{4}+\bs{x}_{2}-\bs{x}_{1})}_\text{type II} \nonumber \\
 & +\underbrace{C_{0}^{(2)}(\bs{r}-\bs{s}+\bs{x}_{3}-\bs{x}_{4})C_{0}^{(2)}(\bs{r}-\bs{s}+\bs{x}_{2}-\bs{x}_{1})}_\text{type II} 
 \nonumber \\&
 +\underbrace{C_{0,c}^{(4)}(\bs{0},\bs{x}_{1}-\bs{x}_{2},\bs{r}-\bs{s},\bs{r}-\bs{s}+\bs{x}_{3}-\bs{x}_{4})}_\text{type III}.\label{C04a}
\end{align}
As before we omitted terms involving $\langle\phi(\bs{x})\rangle_0=\phi_0$ since the latter is assumed to be zero, which therefore means that the connected two-point function is $C_{0,c}^{(2)} = C_{0}^{(2)}$. 

We distinguish, as indicated in (\ref{C04a}), three types of terms:
\begin{itemize}
\item type I is a product of two-point functions $C_{0}^{(2)}$ whose arguments do not depend on the integration variables $\bs{s}$ and $\bs{r}$, 
\item type II are all other terms that are products of two-point functions $C_{0}^{(2)}$ too, but do depend on the integration variables,
\item type III is the connected four-point function $C_{0,c}^{(4)}$.
\end{itemize}

The cluster decomposition principle requires that the connected correlation functions $C_{0,c}^{(2)} = C_{0}^{(2)}$ and $C_{0,c}^{(4)}$ decay to zero when the distance between their points increases. We then readily see that, substituting (\ref{C04a}) into (\ref{eq:6a}) and performing the integration over $\bs{s}$ and $\bs{r}$, only the type I term gives a contribution of order $L^{2d}$ as required in order to survive in the thermodynamic limit, while all the rest give contributions of lower order in $L$.

Similarly, for the $\ddddot{C}_{0}^{(4)}$ correlation function we have
\begin{align}
 & \ddddot{C}_{0}^{(4)}\left(\bs{s},\bs{s}'+\bs{s}+\bs{x}_{1}-\bs{x}_{2},\bs{r},\bs{r}'+\bs{r}+\bs{x}_{3}-\bs{x}_{4}\right)= \nonumber \\
 & = \underbrace{\ddot{C}_{0}^{(2)}\left(\bs{x}_{1}-\bs{x}_{2}+\bs{s}'\right)\ddot{C}_{0}^{(2)}\left(\bs{x}_{3}-\bs{x}_{4}+\bs{r}'\right)}_\text{type I} \nonumber \\ &
 + \underbrace{\ddot{C}_{0}^{(2)}\left(\bs{r}-\bs{s}\right)\ddot{C}_{0}^{(2)}\left(\bs{r}'-\bs{s}'+\bs{r}-\bs{s}+\bs{x}_{3}-\bs{x}_{4}+\bs{x}_{2}-\bs{x}_{1}\right) }_\text{type II} \nonumber \\ &
 +\underbrace{\ddot{C}_{0}^{(2)}\left(\bs{r}'+\bs{r}-\bs{s}+\bs{x}_{3}-\bs{x}_{4}\right)\ddot{C}_{0}^{(2)}\left(\bs{r}-\bs{s}-\bs{s}'+\bs{x}_{2}-\bs{x}_{1}\right) }_\text{type II}\nonumber \\ &
 +\underbrace{\ddddot{C_{}}_{0,c}^{(4)}\left(\bs{0},\bs{x}_{1}-\bs{x}_{2}+\bs{s}',\bs{r}-\bs{s},\bs{r}'+\bs{r}-\bs{s}+\bs{x}_{3}-\bs{x}_{4}\right)}_\text{type III}, 
 \label{D04a}
\end{align}
and analogously for $\ddot{C}_{0}^{(4)}$. In the above, the type I term depends only on $\bs{s}'$ and $\bs{r}'$ but not on $\bs{s}$ and $\bs{r}$ as the other terms. 

If the post-quench hamiltonian is massive, then similar conclusions to the ones after (\ref{C04a}) hold for the contribution of the above terms in the thermodynamic limit after all integrations have been performed. Indeed, in this case the function $H(r)$, given by (\ref{eq:Ha}), decays exponentially with the distance, therefore once again only the type I term gives a contribution that scales as $L^{2d}$. 

Summing up all surviving terms, (\ref{eq:6a}) reduces to
\begin{align}
& \bar{C}^{(4)}(\bs{x}_{1},\bs{x}_{2},\bs{x}_{3},\bs{x}_{4}) =\frac{1}{32} \sum_{{\text{all perm.s}\atop \text{of }1,2,3,4}} \nonumber \\
 & \left(C_{0}^{(2)}(\bs{x}_{1}-\bs{x}_{2})+\int d^ds'H(\bs{s}')\ddot{C}_{0}^{(2)}(\bs{x}_{1}-\bs{x}_{2}+\bs{s}')\right) \nonumber \\
 & \times\left(C_{0}^{(2)}(\bs{x}_{3}-\bs{x}_{4})+\int d^dr'H(\bs{r}')\ddot{C}_{0}^{(2)}(\bs{x}_{3}-\bs{x}_{4}+\bs{r}')\right), \label{app:eq:C4a}
\end{align}
which, using (\ref{eq:Cinfty2}) for the two-point correlation
function, can be written as
\begin{align}
& \bar{C}^{(4)}(\bs{x}_{1},\bs{x}_{2},\bs{x}_{3},\bs{x}_{4}) \nonumber \\
 & = \frac{1}{8}\sum_{{\text{all perm.s}\atop \text{of }1,2,3,4}} \bar{C}^{(2)}(\bs{x}_{1}-\bs{x}_{2})\bar{C}^{(2)}(\bs{x}_{3}-\bs{x}_{4}) \nonumber \\ 
& =\bar{C}^{(2)}(\bs{x}_{1}-\bs{x}_{2})\bar{C}^{(2)}(\bs{x}_{3}-\bs{x}_{4})+[2\leftrightarrow3]+[2\leftrightarrow4],\label{app:eq:C4ba}
\end{align}
This is identical to Wick's theorem expansion for the four-point function.

If, on the other hand, the post-quench hamiltonian is massless, then $H(r)$ does not decay exponentially and therefore it is no longer certain that the asymptotic expressions of the integrals scale with $L$ as they should in order to give a finite (non-divergent) contribution in the thermodynamic limit. More specifically, if $H(r)$ diverges with $L$ (as in $d=2$) then the same is certainly true for the long time averaged correlation functions, which therefore do not equilibrate. Even if $H(r)$ does not diverge with $L$ but decays algebraically with the distance as $r^{-2}$ or slower, then it is not an integrable weight function and the long time averaged functions may diverge with $L$, depending on the large distance behaviour of the initial correlation functions. However it is still true that the dominant contribution in the thermodynamic limit is given only by the type I terms above, since the cluster decomposition principle ensures that the spatial averages of all other types of terms scale slower with $L$. Therefore (\ref{app:eq:C4a}) above is still applicable at leading order in $L$. We denote this by replacing the equality symbol by $ \overset{L\to \infty}{\longrightarrow} $, meaning that only the leading asymptotic expression is kept, while the omitted subdominant terms may also diverge with $L$ but slower. Possible exceptions to this conclusion may arise only in cases where the initial correlation functions decay algebraically with exponents that have special values, such that type III terms do contribute to the leading-order in the thermodynamic limit.

\section{The GGE predictions for the correlation functions}
\label{app:3}

We will calculate the GGE prediction for the two-point correlation
function. The GGE is given, always in the thermodynamic limit, by
the (non-normalised) density matrix 
\begin{equation}
\rho_{\rm GGE}={\exp{\left(-\int {d^d{k}} \, \beta(\bs{k})n(\bs{k})\right)}},\label{app:gge}
\end{equation}
where the Lagrange multipliers $\beta(\bs{k})$ are defined through the 
requirement that the values of the charges $n(\bs{k})$ in the GGE are
equal to their initial values
\begin{equation}
\langle n(\bs{k})\rangle_{\rm GGE}=\langle n(\bs{k})\rangle_{0}.\label{app:gge3}
\end{equation}
The GGE value of the two-point function is then
\begin{align}
& C_{\text{GGE}}^{(2)}(\bs{x},\bs{y})  \equiv\langle\phi(\bs{x})\phi(\bs{y})\rangle_{\text{GGE}} \nonumber \\
& =\int\frac{d^d{k}}{(2\pi)^d}\frac{1}{2E_{\bs{k}}}e^{i\bs{k}\cdot(\bs{x}-\bs{y})}\ensuremath{\left(\langle a_{-\bs{k}}^{\dagger}a_{-\bs{k}}\rangle_{\text{GGE}}+\langle a_{\bs{k}}a_{\bs{k}}^{\dagger}\rangle_{\text{GGE}}\right)} \nonumber \\
& =\int\frac{d^d{k}}{(2\pi)^d}\frac{1}{2E_{\bs{k}}}e^{i\bs{k}\cdot(\bs{x}-\bs{y})}\ensuremath{\left(1+\langle n(-\bs{k})\rangle_{\text{GGE}}+\langle n(\bs{k})\rangle_{\text{GGE}}\right)},
 \label{app:2pt_gge}
\end{align}


Since the GGE, as expressed in (\ref{gge2}), is diagonal in the momentum basis, it is a gaussian ensemble and therefore Wick's theorem for the expectation values of multi-point correlation functions is valid in it. This means that the prediction for the four-point function is the sum of all possible disconnected two-point function contractions, that
is 
\begin{align}
 & C_{\text{GGE}}^{(4)}(\bs{x}_{1},\bs{x}_{2},\bs{x}_{3},\bs{x}_{4}) \nonumber \\
& = C_{\text{GGE}}^{(2)}(\bs{x}_{1},\bs{x}_{2})C_{\text{GGE}}^{(2)}(\bs{x}_{3},\bs{x}_{4})+[2\leftrightarrow3]+[2\leftrightarrow4]\label{app:eq:G4gge}\\
 & =\int\frac{d^dk\, d^d{p}}{(2\pi)^{2d}}\frac{1}{4E_{\bs{k}}E_{\bs{p}}}e^{i\bs{k}\cdot(\bs{x}_{1}-\bs{x}_{2})+i\bs{p}\cdot(\bs{x}_{3}-\bs{x}_{4})} \nonumber \\
 & \times {\left(\langle n(-\bs{k})\rangle_{\text{GGE}}+\langle n(\bs{k})\rangle_{\text{GGE}}+1\right)} \nonumber \\
 & \times {\left(\langle n(-\bs{p})\rangle_{\text{GGE}}+\langle n(\bs{p})\rangle_{\text{GGE}}+1\right)} \nonumber \\
 & + [2\leftrightarrow3]+[2\leftrightarrow4].\nonumber 
\end{align}


\begin{thebibliography}{10}
\bibitem{revq}
A. Polkovnikov, K. Sengupta, A. Silva, and M. Vengalattore, Rev. Mod. Phys. {\bf 83}, 863 (2011).

\bibitem{revq1}
V.I. Yukalov, Laser Phys. Lett. {\bf8} (2011) 485-507.

\bibitem{revq2}
J. Eisert, M. Friesdorf, C. Gogolin, Nat. Phys. {\bf11}, 124 (2015); \\
C. Gogolin, J. Eisert, arXiv:1503.07538 

\bibitem{cc06} 
P. Calabrese and  J. Cardy, Phys. Rev. Lett. {\bf 96}, 136801 (2006); \\
J. Stat. Mech. P06008  (2007); \\
J. Stat. Mech. P04010 (2005).

\bibitem{CEF} 
P. Calabrese, F.H.L. Essler and M. Fagotti, Phys. Rev. Lett. \textbf{106}, 227203 (2011); \\
J. Stat. Mech.  P07016 (2012); \\
J. Stat. Mech. P07022 (2012).

\bibitem{rdyo07} 
M. Rigol, V. Dunjko, V. Yurovsky,  and M. Olshanii, Phys. Rev. Lett. {\bf 98},0 50405 (2007).

\bibitem{uc}
M.~Greiner, O.~Mandel, T.~W.~H\"ansch, and I.~Bloch,
Nature {\bf 419} 51 (2002).

\bibitem{kww06}
T. Kinoshita, T. Wenger,  D. S. Weiss,  Nature {\bf 440}, 900 (2006).

\bibitem{tc07}
S. Hofferberth, I. Lesanovsky, B. Fischer, T. Schumm, and J. Schmiedmayer,
Nature {\bf 449}, 324 (2007).

\bibitem{tetal11}
S. Trotzky Y.-A. Chen, A. Flesch, I. P. McCulloch, U. Schollw\"ock,
J. Eisert, and I. Bloch, 
Nature Phys. {\bf 8}, 325 (2012). 

\bibitem{cetal12}
M. Cheneau, P. Barmettler, D. Poletti, M. Endres, P. Schauss, T. Fukuhara, C. Gross, I. Bloch, C. Kollath, and S. Kuhr,
Nature {\bf 481}, 484 (2012).

\bibitem{getal11}
M. Gring, M. Kuhnert, T. Langen, T. Kitagawa, B. Rauer, M. Schreitl, I. Mazets, D. A. Smith, E. Demler, and J. Schmiedmayer,
Science {\bf 337}, 1318 (2012).

\bibitem{shr12}
U. Schneider, L. Hackerm\"uller, J. P. Ronzheimer, S. Will, S. Braun, T. Best, I. Bloch, E. Demler, S. Mandt, D. Rasch, and A. Rosch,
Nature Phys. {\bf 8}, 213 (2012).

\bibitem{rsb13}
J. P. Ronzheimer, M. Schreiber, S. Braun, S. S. Hodgman, S. Langer, I. P. McCulloch, F. Heidrich-Meisner, I. Bloch, and U. Schneider, Phys. Rev. Lett. {\bf 110}, 205301 (2013).

\bibitem{exp2}
T. Schweigler, V. Kasper, S. Erne, B. Rauer, T. Langen, T. Gasenzer, J. Berges, J. Schmiedmayer, arXiv:1505.03126.


\bibitem{exp1}
T. Langen, S. Erne, R. Geiger, B. Rauer, T. Schweigler, M. Kuhnert, W. Rohringer, I. E. Mazets, T. Gasenzer, J. Schmiedmayer, Science 348 (2015) 207-211.

\bibitem{rdyo07b} 
M. Rigol, V. Dunjko,  and M. Olshanii, Nature {\bf 452}, 854 (2008). 

\bibitem{fm-10}
D. Fioretto and G. Mussardo New J. Phys. {\bf 12} 055015 (2010).

\bibitem{SFM}
S. Sotiriadis, D. Fioretto, G. Mussardo, J. Stat. Mech. P02017 (2012).

\bibitem{Mussardo13}
G. Mussardo, Phys. Rev. Lett. {\bf 111}, 100401 (2013).

\bibitem{c06a} M. A. Cazalilla, Phys. Rev. Lett. {\bf 97}, 156403 (2006) 

\bibitem{c06b} A. Iucci, and M. A. Cazalilla, Phys. Rev. A {\bf 80}, 063619 (2009) 

\bibitem{c06c} A. Iucci, and M. A. Cazalilla, New J. Phys. {\bf 12}, 055019 (2010).

\bibitem{c06}
A. Mitra and T. Giamarchi, Phys. Rev. Lett. {\bf 107}, 150602 (2011).

\bibitem{ir-11}
F. Igloi and H. Rieger,   Phys. Rev. Lett. {\bf 106}, 035701 (2011); \\
Phys. Rev. B. {\bf 84}, 165117 (2011).

\bibitem{cdeo08}
M. Cramer, C. M. Dawson, J. Eisert, and T. J. Osborne, Phys. Rev. Lett. {\bf 100}, 030602 (2008).

\bibitem{cra10}
M. Cramer and J. Eisert, New J. Phys. {\bf 12}, 055020 (2010).

\bibitem{bs08}
T. Barthel and U. Schollw\"ock, Phys. Rev. Lett. {\bf 100}, 100601 (2008).

\bibitem{mgs}
M. Marcuzzi, J. Marino, A. Gambassi, and A. Silva, Phys. Rev. Lett. {\bf111}, 197203 (2013).

\bibitem{ce13}
J.-S. Caux, F. Essler, Phys. Rev. Lett. {\bf 110} 257203 (2013).


\bibitem{fe13}
M.  Fagotti and  F. H. L. Essler, Phys. Rev. B {\bf 87}, 245107 (2013).


\bibitem{fcec13}
M. Fagotti, M. Collura, F. H. L. Essler and P. Calabrese, Phys. Rev. B {\bf 89}, 125101 (2014).


\bibitem{gge-new3}
M. Kormos, A. Shashi, Y.-Z. Chou, J.-S. Caux, A. Imambekov, Phys. Rev. B \textbf{88}, 205131 (2013).

\bibitem{SotiriadisCalabrese}
 S. Sotiriadis, P. Calabrese, J. Stat. Mech. (2014) P07024


\bibitem{bse-14}
B. Bertini, D. Schuricht, F. H. L. Essler, J. Stat. Mech. P10035 (2014).

\bibitem{delf14}
G. Delfino, J. Phys. A: Math. Theor. 47 (2014) 402001.

\bibitem{rig09}
M. Rigol, Phys. Rev. Lett. \textbf{103}, 100403 (2009).

\bibitem{deu}
J. M. Deutsch, Phys. Rev. A {\bf43}, 2046 (1991).

\bibitem{gme}
C. Gogolin, M. P. Muller, and J. Eisert, Phys. Rev. Lett. {\bf106}, 040401 (2011).

\bibitem{rigsre}
M. Rigol and M. Srednicki, Phys. Rev. Lett. {\bf108}, 110601 (2012).

\bibitem{sre}
M. Srednicki, Phys. Rev. E {\bf50}, 888 (1994); \\
J. Phys. A {\bf29}, L75 (1996); \\
{\bf32}, 1163 (1999).

\bibitem{ekmr-14}
F. H. L. Essler, S. Kehrein, S. R. Manmana, and N. J. Robinson,  Phys. Rev. B {\bf 89}, 165104 (2014).

\bibitem{Jaynes}E. T. Jaynes, Phys. Rev. \textbf{106}, 620 (1957); \\
Phys. Rev. \textbf{108}, 171 (1957).

\bibitem{cc-07} 
P. Calabrese and J. Cardy, J. Stat. Mech. P06008 (2007).

\bibitem{cardy12} J. Cardy, ``Quantum Quenches in Perturbed Conformal Field Theories" (2012), talk at GGI workshop: New quantum states of matter in and out of equilibrium; \\ 
J. Cardy, \href{http://online.kitp.ucsb.edu/online/qdynamics-c12/cardy/}{``Quantum Quench in a Conformal Field Theory From a General Short-Range StateÓ} (2012), talk at KITP Conference: Dynamics and Thermodynamics in Isolated Quantum Systems; \\
J. Cardy, arXiv:1507.07266

\bibitem{bs-08} T. Barthel and U. Schollwoeck, Phys. Rev. Lett. \textbf{100}, 100601 (2008).

\bibitem{scc-09}
S. Sotiriadis, P. Calabrese,  and J. Cardy, EPL {\bf 87}, 20002 (2009).

\bibitem{r-09a}M. Rigol, Phys. Rev. A \textbf{80}, 053607 (2009).

\bibitem{caz}
M. A. Cazalilla, A. Iucci, and M.-C. Chung, Phys. Rev. E {\bf 85}, 011133 (2012). 

\bibitem{f-13} M. Fagotti, Phys. Rev. B \textbf{87}, 165106 (2013).

\bibitem{eef-12} F. H. L. Essler, S. Evangelisti, and M. Fagotti,
Phys. Rev. Lett. \textbf{109}, 247206 (2012).

\bibitem{se-12} D. Schuricht and F. H. L. Essler, J. Stat. Mech. P04017 (2012).

\bibitem{ccss-11} T. Caneva, E. Canovi, D. Rossini, G. E. Santoro, and A. Silva, 
J. Stat. Mech.  P07015 (2011).

\bibitem{rs-12} M. Rigol and M. Srednicki, Phys. Rev. Lett. \textbf{108}, 110601 (2012).

\bibitem{CE-08}
M. Cramer, C. M. Dawson, J. Eisert, and T. J. Osborne, 
Phys. Rev. Lett. {\bf100}, 030602 (2008).

\bibitem{CE-10}
M. Cramer and J. Eisert,
New J. Phys. {\bf 12}, 055020 (2010).

\bibitem{CSC13a} M. Collura, S. Sotiriadis and P. Calabrese, Phys. Rev. Lett. \textbf{110}, 245301 (2013); \\
J. Stat. Mech. P09025 (2013).

\bibitem{fe-13} M. Fagotti and F. H. L. Essler, Phys. Rev. B \textbf{87}, 245107 (2013).

\bibitem{RS13} M. A. Rajabpour and S. Sotiriadis, Phys. Rev. A \textbf{89}, 033620 (2014).

\bibitem{Mossel} J. Mossel and J. S. Caux, New J. Phys. \textbf{14} 075006 (2012).

\bibitem{Pozsgay11}B. Pozsgay, J. Stat. Mech.  P01011 (2011).

\bibitem{ce-13} J.-S. Caux and F. H. L. Essler, Phys. Rev. Lett. \textbf{110}, 257203 (2013).

\bibitem{KCC-13}
M. Kormos, M. Collura and P. Calabrese, Phys. Rev. A {\bf 89}, 013609 (2014).

\bibitem{f-14}
M. Fagotti, J. Stat. Mech. P03016 (2014).

\bibitem{ck-14}
M. Collura and D. Karevski, Phys. Rev. B {\bf89}, 214308 (2014).

\bibitem{DMV}
A. De Luca, G. Martelloni and J. Viti, Phys. Rev. A {\bf91}, 021603(R) (2015).

\bibitem{GGE-CFT}
G. Mandal, R. Sinha, and N. Sorokhaibam,
arXiv:1501.04580 [hep-th].

\bibitem{WDNBFRC}
B. Wouters, J. De Nardis, M. Brockmann, D. Fioretto, M. Rigol, and J.-S. Caux,
Phys. Rev. Lett. \textbf{113}, 117202 (2014)

\bibitem{BWFDNVC}
M. Brockmann, B. Wouters, D. Fioretto, J. De Nardis, R. Vlijm and J.-S. Caux,
J. Stat. Mech. (2014) P12009

\bibitem{PMWKZT}
B. Pozsgay, M. Mestyan, M. A. Werner, M. Kormos, G. Zarand, and G. Takacs,
Phys. Rev. Lett. \textbf{113}, 117203 (2014)

\bibitem{POZ1}
B. Pozsgay, arXiv:1406.4613 (2014)

\bibitem{POZ2}
M. Mestyan, B. Pozsgay, G. Takacs, M.A. Werner, arXiv:1412.4787 (2014)

\bibitem{ANDREI}
G. Goldstein, N. Andrei, arXiv:1405.4224 (2014)

\bibitem{PROSEN}
M, Mierzejewski, P. Prelovsek, T. Prosen, Phys. Rev. Lett. {\bf113} 020602 (2014)

\bibitem{Prosen2} T. Prosen, Phys. Rev. Lett. {\bf106}, 217206 (2011). 

\bibitem{Prosen3} T. Prosen, Nuclear Physics B {\bf886}, 1177 (2014).

\bibitem{Pereira}
R. G. Pereira, V. Pasquier, J. Sirker and I. Affleck, J. Stat. Mech. (2014) P09037.

\bibitem{Panfil}
F.H.L. Essler, G. Mussardo, M. Panfil, Phys. Rev. A {\bf91}, 051602 (2015).

\bibitem{CM2014}
M. Collura and G. Martelloni, J. Stat. Mech. P08006 (2014).

\bibitem{BOSONICDOYON}
B. Doyon, A. Lucas, K. Schalm, M. J. Bhaseen, arXiv:1409.6660 (2014)

\bibitem{Hartnoll}
S. A. Hartnoll, Class. Quant. Grav. {\bf26}, 224002 (2009).

\bibitem{MG}
J. McGreevy, 
Advances in High Energy Physics, 723105 (2010).

\bibitem{myers}
S. R. Das, D. A. Galante, R. C. Myers, 
Phys. Rev. Lett. {\bf112}, 171601 (2014); \\
JHEP {\bf02} (2015) 167; \\
arXiv:1505.05224 [hep-th] (2015).

\bibitem{HolographicDoyon}
M. J. Bhaseen, B. Doyon, A. Lucas, and K. Schalm, Nat. Phys. {\bf11}, aop (2015).

\bibitem{AMADO}
I. Amado, A. Yarom, 	arXiv:1501.01627 (2015)

\bibitem{Coleman-Mandula}
S. Coleman, J. Mandula, Phys. Rev. {\bf159}, 1251 (1967).

\bibitem{CM2}
R. Haag, J. T. Lopuszanski, M. Sohnius, Nucl. Phys. B {\bf88} 257 (1975).

\bibitem{CMgen}
J. Maldacena, Al. Zhiboedov 2013 J. Phys. A: Math. Theor. {\bf46} 214011.

\bibitem{MALDACENA1}
  J.~M.~Maldacena,    Int. J. Theor. Phys.  {\bf 38} (1999) 1113; \\
  Adv. Theor. Math. Phys.  {\bf 2} (1998) 231.
      

\bibitem{Beisert:2010jr}
  N.~Beisert, C.~Ahn, L.~F.~Alday, Z.~Bajnok, J.~M.~Drummond, L.~Freyhult, N.~Gromov and R.~A.~Janik {\it et al.},  Lett. Math. Phys.  {\bf 99} (2012) 3.

\bibitem{MALDACENA2}
O.~Aharony, O.~Bergman, D.~L.~Jafferis and J.~Maldacena,  JHEP {\bf 0810} (2008) 091
      
      
\bibitem{NIC14} N. Nessi, A. Iucci, and M.?A. Cazalilla, Phys. Rev. Lett. {\bf113}, 210402 (2014)

\bibitem{MSF14} A. Maraga, A. Silva, and M. Fabrizio, Phys. Rev. B {\bf90}, 155131 (2014).

\bibitem{SS8} M. A. Rajabpour, S. Sotiriadis, Phys. Rev. A {\bf89}, 033620 (2014).

\bibitem{SS12} M. A. Rajabpour, S. Sotiriadis, Phys. Rev. B {\bf91}, 045131 (2015).


\bibitem{SC} 
S. Sotiriadis and J. Cardy, Phys. Rev. B \textbf{81}, 134305 (2010).


\bibitem{Drummond:2009fd}
  J.~M.~Drummond, J.~M.~Henn and J.~Plefka,
    JHEP {\bf 0905} (2009) 046.

\bibitem{Gaiotto:2008sa}
  D.~Gaiotto and E.~Witten,
    J.\ Statist.\ Phys.\  {\bf 135} (2009) 789.

\bibitem{Pestun:2007rz}
  V.~Pestun,
      Commun.\ Math.\ Phys.\  {\bf 313} (2012) 71.

\bibitem{qe-hol} 
V. E. Hubeny, M. Rangamani, and T. Takayanagi,
JHEP {\bf 0707} (2007) 062.

\bibitem{qe-hol1} 
J. Abajo-Arrastia, J. Aparicio, and E. Lopez,
JHEP {\bf 1011} (2010) 149.

\bibitem{qe-hol2} 
J. Aparicio and E. Lopez,
JHEP 1112 (2011) 082.

\bibitem{qe-hol3}
V. Balasubramanian, A. Bernamonti, N. Copland, B. Craps, and F. Galli,
Phys. Rev. D {\bf 84} (2011) 105017.

\bibitem{qe-hol4} 
A. Allais and E. Tonni, JHEP {\bf 1201} (2012) 102. 

\bibitem{qe-hol5} 
R. Callan, J. Y. He, and M. Headrick,
JHEP {\bf 1206} (2012) 081.

\bibitem{qe-hol6} 
T. Hartman and J. Maldacena,
JHEP {\bf 1305} (2013) 014.

\bibitem{qe-hol7} 
H. Liu and S. J. Suh, 
Phys. Rev. D {\bf 89} (2014) 066012; \\
H. Liu and S. J. Suh,
Phys. Rev. Lett. {\bf 112} (2014) 011601.

\bibitem{G1}
G. Mandal and T. Morita, JHEP 10 (2013) 197.

\bibitem{G2}
P. Caputa, G. Mandal, and R. Sinha, JHEP 1311 (2013) 052.


\bibitem{prethrm1} G. Biroli, C. Kollath, and A.M. Lauchli, Phys. Rev. Lett. \textbf{105}, 250401 (2010).
\bibitem{prethrm2} S. R. Manmana, S. Wessel, R. M. Noack, and A. Muramatsu, Phys. Rev. Lett. \textbf{98} 210405 (2007); Phys. Rev. B \textbf{79}, 155104 (2009); 
C. Kollath, A. M. Lauchli, and E. Altman; Phys. Rev. Lett. \textbf{98}, 180601 (2007).
\bibitem{prethrm3} M. Moeckel and S. Kehrein, Phys. Rev. Lett. \textbf{100}, 175702 (2008); Ann. Phys. \textbf{324}, 2146 (2009); New J. Phys. \textbf{12}, 055016 (2010).
\bibitem{prethrm4} M. Kollar and M. Eckstein,  Phys. Rev. A \textbf{78}, 013626 (2008); M. Eckstein, M. Kollar, and P. Werner, Phys. Rev. Lett. {\bf103}, 056403 (2009); M. Kollar, F. A. Wolf, and M. Eckstein, Phys. Rev. B \textbf{84}, 054304 (2011).
\bibitem{prethrm5}  L. F. Santos and M. Rigol, Phys. Rev. E \textbf{81}, 036206 (2010); L. F. Santos and M. Rigol, Phys. Rev. E \textbf{82}, 031130 (2010).
\bibitem{prethrm6} G. Brandino, J.-S. Caux, and R. M. Konik, arXiv:1301.0308.

\bibitem{prethrm9} M. van den Worm, B. C. Sawyer, J. J. Bollinger, and M. Kastner, New J. Phys. \textbf{15}, 083007 (2013).

\bibitem{prethrm10} M. Fagotti, J. Stat. Mech. (2014) P03016; B. Bertini, M. Fagotti, J. Stat. Mech. (2015) P07012; M. Fagotti, M. Collura, arxiv:1507.02678.


\end{thebibliography}
\end{document}